\documentclass[automark,headsepline,11pt,oneside,listsleft,DIV12]{scrartcl}
\usepackage{scrpage2}
\clearscrheadfoot
\ohead[]{\pagemark}
\ihead[]{\upshape Agents for Traffic Simulation (A. Kesting, M. Treiber and D. Helbing)}
\usepackage{caption2}

\setcaptionwidth{1.0 \textwidth}

\usepackage{multirow}
\usepackage{graphicx}
\usepackage{amssymb,amsmath} 
\usepackage{booktabs,units} 
\providecommand{\ableitung}[2]{\frac{\text{d} #1}{\text{d} #2}}
\newcommand{\subpath}{.}
\newcommand{\cpath}[1]{\subpath/#1}
\title{Agents for Traffic Simulation}

\author{Arne Kesting$^1$, Martin Treiber$^1$ and Dirk Helbing$^2$\\[1ex]
$^{1}$Technische Universit\"at Dresden\\
Institute for Transport \& Economics\\
Andreas-Schubert-Str.~23, 01062 Dresden (Germany)\\
$^2$ETH Zurich, UNO D11, \\
Universit{\"a}tsstr.~41, CH-8092 Zurich (Switzerland)}
\date{April 14, 2008}
\begin{document}
\maketitle
\footnotetext[1]{E-mail: {\tt kesting@vwi.tu-dresden.de}, URL: {\tt http://www.akesting.de}}
\begin{abstract}
Vehicular traffic is a classical example of a multi-agent system in
which autonomous drivers operate in a shared environment. The article
provides an overview of the state-of-the-art in microscopic traffic
modeling and the implications for simulation techniques. We focus on
the short-time dynamics of car-following models which describe
continuous feedback control tasks (acceleration and braking) and
models for discrete-choice tasks as a response to the surrounding
traffic. The driving style of an agent is characterized by model
parameters such as reaction time, desired speed, desired time gap,
anticipation etc. In addition, internal state variables corresponding
to the agent's ``mind'' are used to incorporate the driving
experiences. We introduce a time-dependency of some parameters to
describe the frustration of drivers being in a traffic jam for a
while. Furthermore, the driver's behavior is externally influenced by
the neighboring vehicles and also by environmental input such as
limited motorization and braking power, visibility conditions and road
traffic regulations. A general approach for dealing with discrete
decision problems in the context of vehicular traffic is introduced
and applied to mandatory and discretionary lane changes. Furthermore,
we consider the decision process whether to brake or not when
approaching a traffic light turning from green to amber. Another
aspect of vehicular traffic is related to the heterogeneity of
drivers. We discuss a hybrid system of coupled vehicle and information
flow which can be used for developing and testing applications of
upcoming inter-vehicle communication techniques.
\end{abstract}
\section{Introduction}\label{sec:intro}
\frenchspacing
\noindent 
\looseness-1
Efficient transportation systems are essential to the functioning and
prosperity of modern, industrialized societies. Mobility is also an
integral part of our quality of life, sense of self-fulfillment and
personal freedom. Our traffic demands of today are predominantly
served by individual motor vehicle travel which is the primary means
of transportation. However, the limited road capacity and ensuing traffic
congestion has become a severe problem in many countries. Nowadays, we
additionally have to balance the human desire for personal mobility
with the societal concerns about its environmental impact and energy
consumption.  On the one hand, traffic demand can only be affected
indirectly by means of policy measures. On the other hand, an
extension of transport infrastructure is no longer an appropriate or
desirable option in densely populated areas. Moreover, construction
requires high investments and maintenance is costly in the long
run. Therefore, engineers are now seeking solutions to the questions
of how the capacity of the road network could be used more efficiently
and how operations can be improved by way of intelligent
transportation systems (ITS).


In the presence of increasing computing power, realistic microscopic
traffic simulations are becoming a more and more important tool for
diverse purposes ranging from generating surrounding traffic in a
virtual reality driving simulator to large-scale network simulations
for a model-based prediction of travel times and traffic
conditions~\cite{autobahn-nrw-de}. The primary application for traffic
simulations is the evaluation of hypothetical scenarios for their
impact on traffic. Computer simulations can be valuable in making
these analyses in a cost-effective way. For example, simulations can
be used to estimate the impact of future driver assistance systems and
wireless communication technologies on traffic dynamics. Another
example is the prediction of congestion levels in the future, based on
demographic forecasts.

Before going into detail about possible traffic flow models, it is
worth mentioning differences between modeling the short-term traffic
dynamics on a single road section and the approach used for
transportation planning describing behavioral pattern in a network on
a larger time scale. Figure~\ref{tab:Timescales} shows typical time
scales ranging over nine orders of magnitude including vehicle
dynamics, traffic dynamics and transportation planning. While dynamic
flow models explicitly describe the \textit{physical propagation of
traffic flows} of a given traffic volume in a road network,
transportation planning tools deal with the calculation of the traffic
demand by considering the \textit{decisions of travelers} to
participate in economical, social and cultural activities. The need
for transportation arises because these activities are spatially
separated. The classical approach in trip-based transportation models
is based on a four-step methodology of \textit{trip generation}, {\it
trip distribution}, {\it mode split}\index{mode split} and {\it
traffic
assignment}~\cite{ortuzar-book,SchnabelLohse,daganzo-book1,Maerivoet-models,helbingNagel2004}. In
the fourth step, the origin-destination matrix of trips with a typical
minimum disaggregation of one hour (comprising a typical peak-hour
analysis) is assigned to routes in the actual (or prospective)
transportation network while taking into account the limited capacity
of the road infrastructure by means of simplified effective
models. Recently, even large-scale multi-agent transportation
simulations have been performed in which each traveler is represented
individually~\cite{nagel-2000-lst,Raney-2003-swiss,charyparNagel2005}. For
the purposes of demand-modeling, mobility-simulation and
infrastructure re-planning the open-source software \textsc{MATSim}
provides a toolbox to implement large-scale agent-based transport
simulations~\cite{Matsim}.

\begin{table}
\centering
{\small
\begin{tabular}{rlll} 
\toprule
Time Scale & Subject & Models & Aspects \\
\midrule
$\unit[0.1]{s}$    & Vehicle Dynamics  & Sub-microscopic & Drive-train, brake, ESP \\
$\unit[1]{s}$      & \multirow{4}{*}{Traffic Dynamics} & \multirow{4}{*}{\parbox{45mm}{Car-following models\\Fluid-dynamic models}} & Reaction time, Time gap \\
$\unit[10]{s}$     & & & Accelerating and braking   \\
$\unit[1]{min}$    & & & Traffic light period \\
$\unit[10]{min}$   & & & Period of stop-and-go wave\\
$\unit[1]{h}$      & \multirow{5}{*}{Transport. Planning} & \multirow{3}{*}{\parbox{45mm}{Traffic assignment models\\ Traffic demand model}} & Peak hour\\
$\unit[1]{day}$    & & & Day-to-day human behavior \\
$\unit[1]{year}$   & & & Building measures \\
$\unit[5]{years}$  & & Statistics & Changes in spatial structure\\
$\unit[50]{years}$ & & Prognosis & Changes in Demography \\
\bottomrule
\end{tabular}
}

 \caption{\label{tab:Timescales}Subjects in transportation systems
 sorted by typical time scales involved. }

\end{table}

\subsection{Aim and Overview}\label{sec:overview}

The chapter will review the state-of-the-art in microscopic traffic
flow modeling and the implications for simulation techniques. In
Sec.~\ref{sec:agent}, we will introduce the concept of a
\textit{driver-vehicle agent} within in the context of common
traffic modeling approaches.

In order to perform traffic simulations, we will take a ``bottom-up''
approach and present concrete models for describing the behavior of an
agent. In Sec.~\ref{sec:IDM}, the Intelligent Driver Model~\cite{Opus}
serves as a basic example of a car-following model representing the
operational level of driving. As a first contribution, we will give
special attention to the heterogeneity in traffic. Different drivers
behave differently in the same situation (so called ``inter-driver
variability'') but can also change their behavior over the course of
time (``intra-driver variability''). While the first aspect can be
addressed by individual parameter sets for the agents
(Sec.~\ref{sec:inter}), the latter can be modeled by introducing a
time-dependency of some parameters (e.g.\ to model the frustration of
drivers after being in a traffic jam for a period, Sec.~\ref{sec:memory}).

Realistic simulations of multi-lane freeway traffic and traffic in
city networks also require discrete decisions by the agents. For
example, lane-changing decisions allow faster cars to pass slower
trucks. Another decision is related to the decision process of whether
to brake or not to brake when approaching a traffic light turning from
green to amber. In Sec.~\ref{sec:MOBIL}, we will introduce a general
framework for dealing with these discrete decision processes. The
presented ``meta-model'' MOBIL~\cite{MOBIL-TRR07} is an example of how
complexity can in fact be reduced by falling back on the agent's model
for calculating longitudinal accelerations.

After having modeled the agent's acceleration and lane-changing
behavior, we will consider multi-agent
simulations. Section~\ref{sec:simulator} addresses the design of
microscopic traffic simulators. In order to be specific, we will
discuss the explicit numerical integration scheme, input and output
quantities and visualization possibilities.

In Sec.~\ref{sec:sim}, we will demonstrate the expressive power of the
agent-based approach for handling current research questions. Traffic
simulations will illustrate the emergence of collective dynamics from
local interaction between agents. By way of example, we will show how
the desired individual behavior of agents to move forward fast can
lead to contrary effects such as the breakdown of traffic and self-organized
stop-and-go waves. Another simulation will evaluate the effect of
traffic flow homogenization by means of a speed limit
(Sec.~\ref{sec:speedlimit}). Last but not least, we will discuss an
application of inter-vehicle communication for propagating
traffic-related information in a decentralized way. Inter-vehicle
communication has recently received  much attention in the academic and
engineering world as it is expected to be a challenging issue for the
next generation of vehicle-based Intelligent Transportation Systems
(ITS). Finally, in Sec.~\ref{sec:outlook}, we will discuss such trends
in traffic modeling and simulation.

\section{Agents for Traffic Simulation}\label{sec:agent}

Vehicular traffic is a typical example of a \textit{multi-agent
system}: Autonomous agents (the drivers) operate in a shared
environment provided by the road infrastructure and react to the
neighboring vehicles.  Therefore, the activities include both
\textit{human interaction} (with the dominant influence originating
from the directly leading vehicle) and
\textit{man-machine-interactions} (driver interaction with the
vehicle and the physical road environment). The \textit{microscopic
modeling} or \textit{agent-based} approach describing the motion of
each individual vehicle has grown in popularity over the last
decade. The following Sec.~\ref{sec:model_approaches} will provide an
overview of common mathematical approaches for describing traffic
dynamics. In Sec.~\ref{sec:micro_approach}, we will introduce the
concept of a ``driver-vehicle agent'' within the context of
microscopic traffic modeling.

\subsection{Macroscopic vs.\ Microscopic Approaches}\label{sec:model_approaches}

The mathematical description of the dynamics of traffic flow has a
long history already. The scientific activity had its beginnings in
the 1930s with the pioneering studies on the fundamental relations of
traffic flow, velocity and density conducted by
Greenshields~\cite{Greenshields}. By the 1950s, scientists had started
to describing the physical propagation of traffic flows by means of
dynamic macroscopic and microscopic models. During the 1990s, the
number of scientists engaged in traffic modeling grew rapidly because
of the availability of better traffic data and higher computational
power for numerical analysis.

Traffic models have been successful in reproducing the observed
\textit{collective}, \textit{self-organized traffic dynamics} including phenomena such
as breakdowns of traffic flow, the propagation of stop-and-go waves
(with a characteristic propagation velocity), the capacity drop, and
different spatiotemporal patterns of congested traffic due to
instabilities and nonlinear
interactions~\cite{Helb-Opus,Kerner-book,Kerner-Rehb96,CasBer-99,Daganzo-ST,Martin-empStates}. For
an overview of experimental studies and the development of
miscellaneous traffic models, we refer to the recently published
extensive review
literature~\cite{Helb-Opus,chowdhury-report,nagatani-report,Maerivoet-report,Hoogendoorn-review,Leutzbach}.

\begin{figure}[t]
\centering
 \includegraphics[width=0.6\linewidth]{\cpath{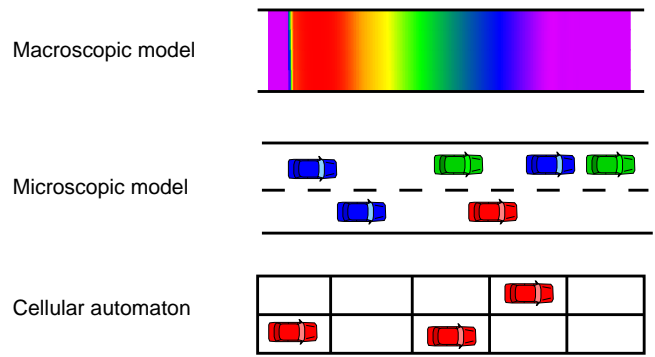}}

 \caption{\label{fig:model_approaches}Illustration of different
 traffic modeling approaches: A snapshot of a road section at time
 $t_0$ is either characterized by {\it macroscopic traffic quantities}
 like traffic density $\rho(x,t_0)$, flow $Q(x,t_0)$ or average
 velocity $V(x,t_0)$, or, {\it microscopically}, by the positions
 $x_\alpha(t_0)$ of single driver-vehicle agent $\alpha$. For cellular
 automata, the road is divided into cells which can be either occupied
 by a vehicle or empty.}

\end{figure}

As mentioned, there are two major approaches to describe the
spatiotemporal propagation of traffic flows. \textit{Macroscopic
traffic flow models} make use of the picture of traffic flow as a
physical flow of a fluid. They describe the traffic dynamics in
terms of aggregated macroscopic quantities like the traffic density,
traffic flow or the average velocity as a function of space and time
corresponding to partial differential equations (cf.\
Fig.~\ref{fig:model_approaches}). The underlying assumption of all
macroscopic models is the conservation of vehicles (expressed by the
continuity equation) which was initially considered by Lighthill,
Whitham and Richards~\cite{Lighthill-W,Richards}. More advanced,
so-called ``second-order'' models additionally treat the macroscopic
velocity as a dynamic variable in order to also consider the finite
acceleration capability of vehicles~\cite{KK-94,GKT}.

By way of contrast, \textit{microscopic traffic models} describe the
motion of each individual vehicle. They model the action such as
accelerations, decelerations and lane changes of each driver as a
response to the surrounding traffic. Microscopic traffic models are
especially suited to the study of heterogeneous traffic streams
consisting of different and individual types of \textit{driver-vehicle
units} or \textit{agents}. The result is individual trajectories
of all vehicles and, consequently, any macroscopic information by
appropriate aggregation. Specifically, one can distinguish the
following major subclasses of microscopic traffic models (cf.\
Fig.~\ref{fig:model_approaches}):
\begin{itemize}
\item
\textit{Time-continuous models} are formulated as ordinary or
delay-differential equations and, consequently, space and time are
treated as continuous variables. \textit{Car-following models} are the
most prominent examples of this approach~\cite{Bando,Opus,Jiang-vDiff01,Tilch-GFM}. In
general, these models are deterministic but stochasticity can be added
in a natural way~\cite{VDT}.  For example, a modified version of the
Wiedemann model \cite{Wiedemann} is used in the commercial traffic simulation software
PTV-VISSIM$^\text{\textsc{\tiny TM}}$.
\item
\textit{Cellular automata} (CA) use
integer variables to describe the dynamic state of the system. The
time is discretized and the road is divided into cells which can be
either occupied by a vehicle or empty. Besides rules for accelerating and
braking, most CA models require additional stochasticity. The first CA
for describing traffic was proposed by Nagel and
Schreckenberg~\cite{Nagel-S}. Although CA lack the accuracy of
time-continuous models, they are able to reproduce some traffic
phenomena~\cite{CA_limitedAcc,Helb-Schreck,Kno01}.  Due to their
simplicity, they can be implemented very efficiently and are suited to
simulating large road networks~\cite{autobahn-nrw-de}.
\item
\textit{Iterated coupled maps} are between CA and time-continuous
models. In this class of model, the update time is considered as an
explicit model parameter rather than an auxiliary parameter needed for
numerical integration~\cite{ThreeTimes-07}. Consequently, the time is
discretized while the spatial coordinate remains continuous. Popular
examples are the Gipps model~\cite{Gipps81} and the Newell
model~\cite{Newell}. However, these models are typically associated
with car-following models as well.
\end{itemize}

At first glance, it may be surprising that simple (and deterministic)
mathematical models aimed at describing the complexity of and
variations in the human behavior, individual skills and driving
attitudes would lead to reasonable results. However, a traffic flow
can (in a good approximation) be considered as a one-dimensional
system (with reduced degrees of freedom). Furthermore, traffic models
typically assume rational and safe driving behavior as a reaction to
the surrounding traffic while at the same time taking into account the
fundamental laws of kinematics.

Another aspect concerns the important issue of traffic safety. The
traditional models for describing traffic dynamics assume rational
drivers that are programmed to avoid collisions.\footnote{Of course,
collisions happen in numerical simulations due to instable models and
for kinematic reasons. However, these collisions do not have
explanatory or predictive power.} Therefore, traffic safety simulation
belongs to the field of human centered simulation where the
perception-reaction system of drivers with all its weak points has to
be described. Up to now, a general modeling approach is still lacking.

\subsection{Driver-Vehicle Agents}\label{sec:micro_approach}

\begin{figure}[t]
\centering
\includegraphics[width=0.8\linewidth]{\cpath{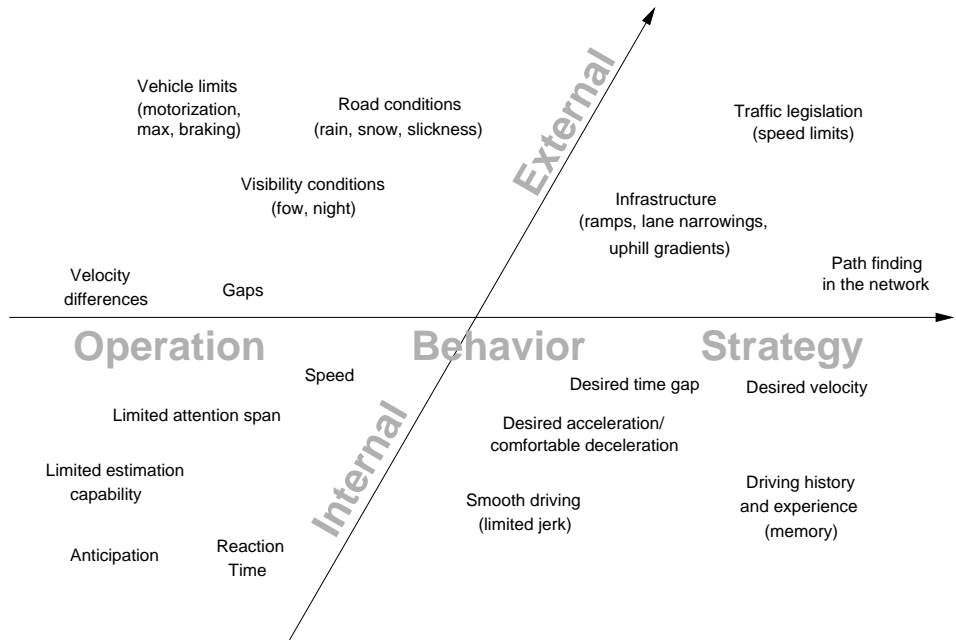}}

 \caption{\label{fig:drivingTask}Characteristics of a driver-vehicle
 agent. The operation of driving can be classified according the
 involved time scales ranging from short-term actions in terms of
 acceleration and braking via intermediate time scales describing
 behavioral characteristics to long-term strategic decisions. In
 addition, the agent's behavior is influenced by the physical
 properties of the vehicle, by interactions with other agents and by
 the environment.}

\end{figure}

Let us now adopt the concept of an agent to implicate the complex
human driving behavior into a general modeling framework. We therefore
introduce the term \textit{driver-vehicle agent} which refers to
the idea that an atomic entity includes internal characteristics of
human drivers as well as external properties of a
vehicle. Figure~\ref{fig:drivingTask} provides an overview of
relevant influences and features affecting human driving
. In this context, the relevant time scales are a first
characteristic feature: The \textit{short-term operations} are
constituted by control tasks such as acceleration and braking, and 
typically take place in the range of a second. Specific
\textit{behavioral attributes} vary between individual drivers and
affect the resulting driving characteristics on an intermediate time
scale of seconds up to minutes. Finally, a strategic level of driving
includes time periods of hours, e.g.\ the decision to start a trip or
to find a route in a network.

The driving task can be considered as a cognitive and therefore
\textit{internal} process of an agent: The driver's perception is limited to
the observable \textit{external} objects in the neighborhood while his
or her reaction is delayed due to a non-negligible reaction time as a
consequence of the physiological aspects of sensing, perceiving,
deciding, and performing an action. On the intermediate time scale,
the agent's actions are affected by his or her individual driving
behavior which may be characterized in terms of, e.g. preferred time
gaps when following a vehicle and smooth driving with a desired
acceleration and a comfortable deceleration. Moreover, the individual
driving style may be influenced by the experience and history of
driving. For example, it is observed that people change their behavior
after being stuck in traffic congestion for a
period~\cite{Brilon-traff95,TGF01}. Such features can be
incorporated by internal state variables corresponding to the agent's
``mind'' or ``memory''.

However, short-time driving operations are mainly direct responses
to the stimulus of the surrounding traffic. The driver's behavior is
externally influenced by environmental input such as limited
motorization and braking power of the vehicle, visibility conditions,
road characteristics such as horizontal curves, lane narrowings,
ramps, gradients and road traffic regulations. In the following
sections, we will address a number of these defining characteristics
of a driver-vehicle agent.

\section{Models for the Driving Task}

Microscopic traffic models describe the motion in longitudinal
direction of each individual vehicle. They model the action of a
driver such as accelerations and decelerations as a response to the
surrounding traffic by means of an {\it acceleration strategy} towards
a desired speed in the free-flow regime, a {\it braking strategy} for
approaching other vehicles or obstacles, and a {\it car-driving
strategy for maintaining a safe distance} when driving behind another
vehicle. Microscopic traffic models typically assume that human
drivers react to the stimulus from neighboring vehicles with the
dominant influence originating from the directly leading vehicle known
as ``follow-the-leader'' or ``car-following'' approximation.

\begin{figure}[th]
 \centering
 \includegraphics[width=90mm]{\cpath{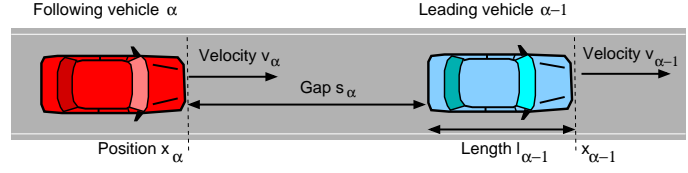}} 

 \caption{\label{fig:IDM_carFollowingSketch}Illustration of the input
 quantities of a car-following model: The bumper-to-bumper distance
 $s$ for a vehicle $\alpha$ with respect to the vehicle $(\alpha-1)$
 in front is given by $s_\alpha = x_{\alpha -1} - x_\alpha -
 l_{\alpha-1}$, where $l_\alpha$ is the vehicle length and $x$ the
 position on the considered road stretch. The approaching rate
 (relative speed) is defined by $\Delta v_\alpha := v_a -
 v_{\alpha-1}$. Notice that the vehicle indices $\alpha$ are ordered
 such that $(\alpha-1)$ denotes the preceding vehicle.}

\end{figure}

By way of example, we will consider the Intelligent Driver Model
(IDM)~\cite{Opus} in Sec.~\ref{sec:IDM}. The IDM belongs to the class
of deterministic follow-the-leader models. Like other car-following
models, the IDM is formulated as an ordinary differential equation
and, consequently, space and time are treated as continuous
variables. This model class is characterized by an acceleration
function $\dot{v}:=\ableitung{v}{t}$ that depends on the actual speed
$v(t)$, the gap $s(t)$ and the velocity difference $\Delta v(t)$ to
the leading vehicle (see Fig.~\ref{fig:IDM_carFollowingSketch}). Note
that the dot is the usual shorthand notation for the time derivative
of a function. The acceleration is therefore defined as the time
derivative of the velocity $\dot{v}:=\text{d}v/\text{d}t$.

In Sec.~\ref{sec:inter}, we will model inter-driver variability by
defining different classes of drivers which is an inherent feature of
microscopic agent approaches. A model for intra-driver variability
(changing behavior over the course of time) will then be discussed in
Sec.~\ref{sec:memory}.

\subsection{The Intelligent Driver Model}\label{sec:IDM}
The IDM acceleration is a continuous function incorporating different
driving modes for all velocities in freeway traffic as well as city
traffic. Besides the distance to the leading vehicle $s$ and the
actual speed $v$, the IDM also takes into account velocity differences
$\Delta v$, which play an essential stabilizing role in real traffic,
especially when approaching traffic jams and avoiding rear-end
collisions (see Fig.~\ref{fig:IDM_carFollowingSketch}). The IDM
acceleration function is given by
\begin{equation}
\label{eq:IDMaccel}
\ableitung{v_{\alpha}}{t} = f(s_\alpha,v_\alpha,\Delta v_\alpha) = a
         \left[ 1 -\left( \frac{v_{\alpha}}{v_0} \right)^\delta -\left(
         \frac{s^*(v_{\alpha},\Delta v_{\alpha})} {s_{\alpha}}
         \right)^2\, \right].
\end{equation}
This expression combines the acceleration strategy
$\dot{v}_\text{free} (v)= a[1-(v/v_0)^\delta]$ towards a
\textit{desired speed} $v_0$ on a free road with the parameter $a$ for
the \textit{maximum acceleration} with a braking strategy
$\dot{v}_\text{brake}(s, v, \Delta v) = -a(s^*/s)^2$ serving as
repulsive interaction when vehicle $\alpha$ comes too close to the
vehicle ahead. If the distance to the leading vehicle, $s_\alpha$, is
large, the interaction term $\dot{v}_\text{brake}$ is negligible and
the IDM equation is reduced to the free-road acceleration
$\dot{v}_\text{free}(v)$, which is a decreasing function of the
velocity with the maximum value $\dot{v}(0)=a$ and the minimum value
$\dot{v}(v_0)=0$ at the desired speed $v_0$. For denser traffic, the
deceleration term becomes relevant. It depends on the ratio between
the effective ``desired minimum gap''
\begin{equation}\label{eq:IDMsstar}
s^*(v, \Delta v)  = s_0  + v T + \frac{v \Delta v }  {2\sqrt{a b}},
\end{equation}
and the actual gap $s_\alpha$. The \textit{minimum distance} $s_0$ in
congested traffic is significant for low velocities only. The main
contribution in stationary traffic is the term $vT$ which corresponds
to following the leading vehicle with a constant {\it desired time
gap} $T$. The last term is only active in non-stationary traffic
corresponding to situations in which $\Delta v\ne 0$ and implements an
``intelligent'' driving behavior including a braking strategy that, in
nearly all situations, limits braking decelerations to the
\textit{comfortable deceleration} $b$. Note, however, that the IDM
brakes stronger than $b$ if the gap becomes too small. This braking
strategy makes the IDM collision-free~\cite{Opus}. All IDM parameters
$v_0$, $T$, $s_0$, $a$ and $b$ are defined by positive values. These
parameters have a reasonable interpretation, are known to be relevant,
are empirically measurable and have realistic
values~\cite{Arne-calibration-TRB}. We will discuss parameter values
in detail in Sec.~\ref{sec:inter} and will use their clear meaning to
characterize different driving styles, that is, inter-driver
variability.

For a simulation scenario with a speed limit (which we will study in
Sec.~\ref{sec:speedlimit}), we consider a refinement of the IDM for
the case when the actual speed is higher than the desired speed,
$v>v_0$. For example, an excess of $v=2 v_0$ would lead to an
unrealistic braking of $-15 a$ for $\delta=4$. This situation may
occur when simulating, e.g.\ a speed limit on a road segment that
reduces the desired speed locally. Therefore, we replace the free
acceleration for the case $v>v_0$ by
\begin{equation}
\dot{v}_\text{free}(v) = -b\left[ 1- \left(\frac{v_0}{v} \right)^\delta  \right].
\end{equation}
That is, the IDM vehicle brakes with the comfortable deceleration $b$
in the limit $v\gg v_0$. Further extensions of the IDM can be found in
Refs.~\cite{HDM,ThreeTimes-07,VDT}.

The \textit{dynamic properties} of the IDM are controlled by the
maximum acceleration $a$, the acceleration exponent $\delta$ and the
parameter for the comfortable braking deceleration $b$. Let us now
consider the following scenario: If the distance $s$ is large
(corresponding to the situation of a nearly empty road), the
interaction $\dot{v}_\text{brake}$ is negligible and the IDM
equation~\eqref{eq:IDMaccel} is reduced to the free-road acceleration
$\dot{v}_\text{free}(v)$. The driver accelerates to his or her desired
speed $v_0$ with the maximum acceleration $\dot{v}(0)=a$. The
acceleration exponent $\delta$ specifies how the acceleration
decreases when approaching the desired speed. The limiting case
$\delta\to\infty$ corresponds to approaching $v_0$ with a constant
acceleration $a$ while $\delta=1$ corresponds to an exponential
relaxation to the desired speed with the relaxation time
$\tau=v_0/a$. In the latter case, the free-traffic acceleration is
equivalent to that of the Optimal Velocity
Model~\cite{Bando}. However, the most realistic behavior is expected
between the two limiting cases of exponential acceleration (for
$\delta = 1$) and constant acceleration (for $\delta \rightarrow
\infty$). Therefore, we set the acceleration exponent constant to
$\delta=4$.

\begin{figure}[t]
\centering

\includegraphics[width=65mm]{\cpath{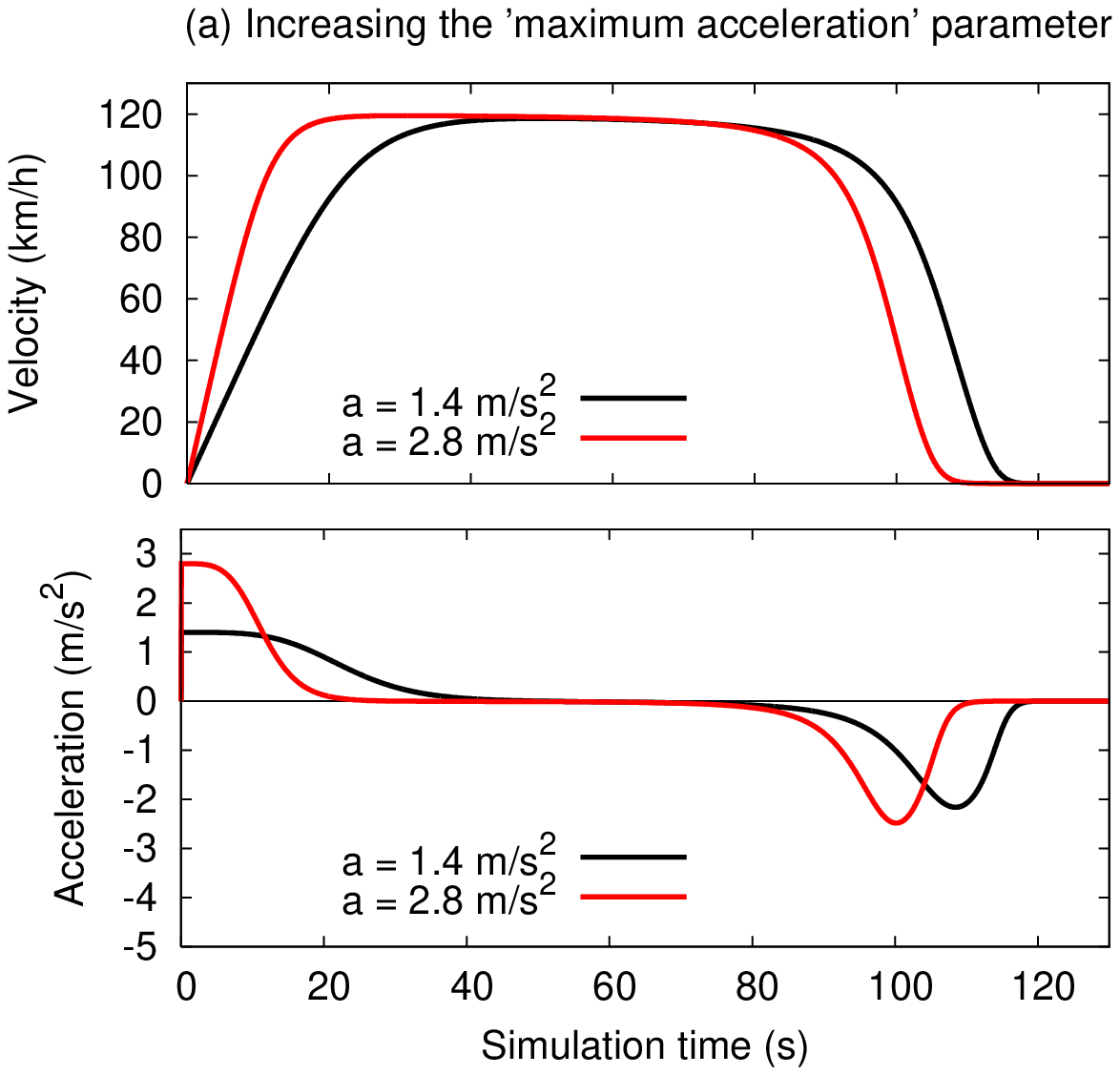}} 
\includegraphics[width=65mm]{\cpath{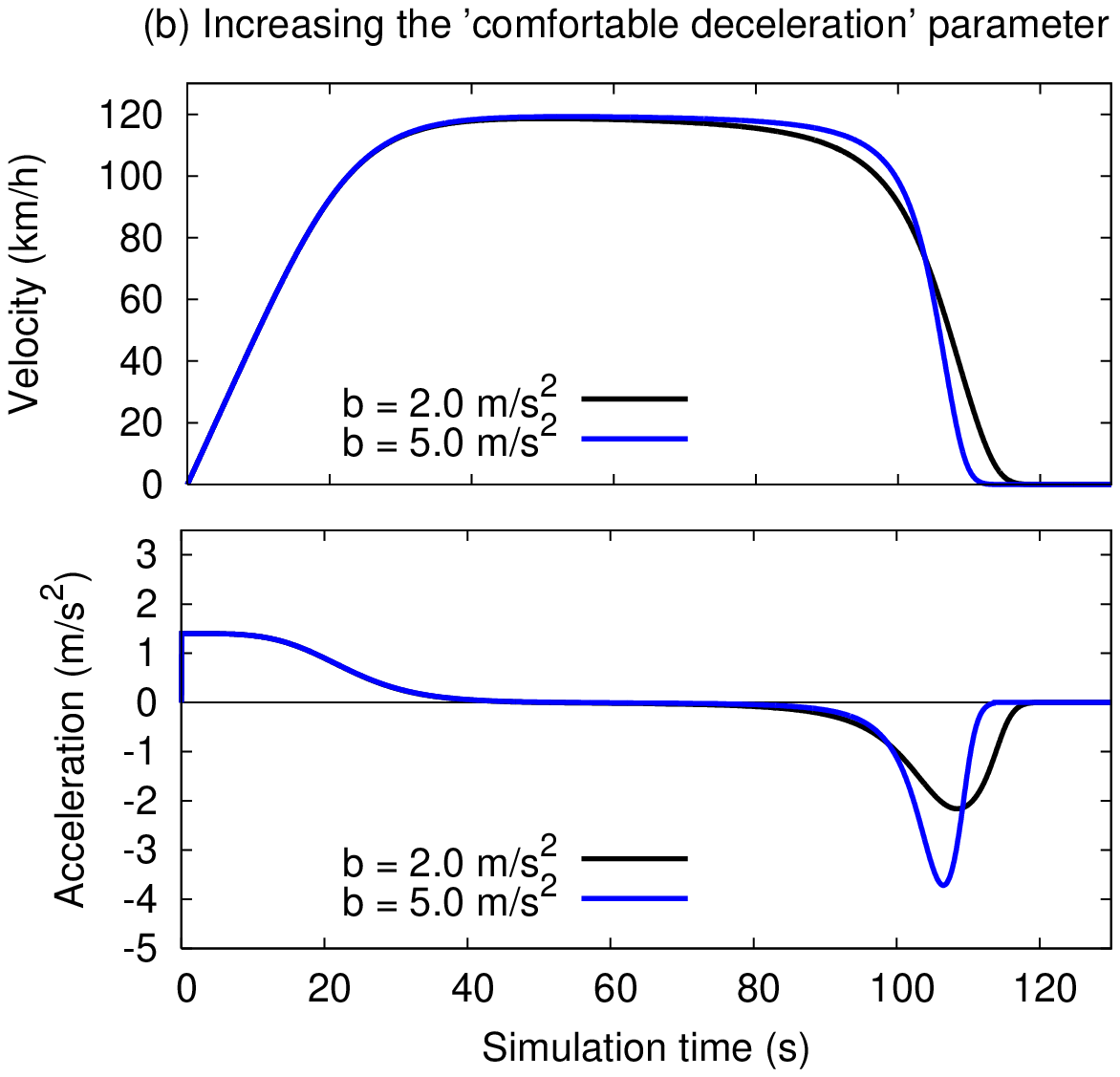}}

  \caption{\label{fig:IDM_single}Simulation of a single driver-vehicle
  agent modeled by the IDM: The diagrams show the acceleration to the
  desired speed $v_0=\unit[120]{km/h}$ followed by braking as a reaction
  to a standing obstacle located \unit[3000]{m} ahead for several
  combinations of the IDM acceleration parameters $a$ [in diagram (a)]
  and $b$ [in (b)]. The remaining parameters are
  $a=\unit[1.4]{m/s^2}$, $b=\unit[2.0]{m/s^2}$, $T=\unit[1.5]{s}$,
  $s_0=\unit[2]{m}$.}

\end{figure}

In Fig.~\ref{fig:IDM_single}, acceleration periods from a standstill
to the desired speed $v_0=120\,$km/h are simulated for two different
settings of the maximum acceleration (the other model parameters are
listed in the caption): For $a=1.4\,$m/s$^2$, the acceleration phase
takes approximately $40\,$s while an increased maximum acceleration of
$a=3\,$m/s$^2$ reduces the acceleration period to $\sim 15\,$s. Notice
that the acceleration parameter $a$ of $1.4\,$m/s$^2$ ($3\,$m/s$^2$)
corresponds to a free-road acceleration from $v=0$ to $v=100\,$km/h
within $23\,$s ($10.5\,$s).

\begin{figure}[t]
\centering
\includegraphics[width=60mm]{\cpath{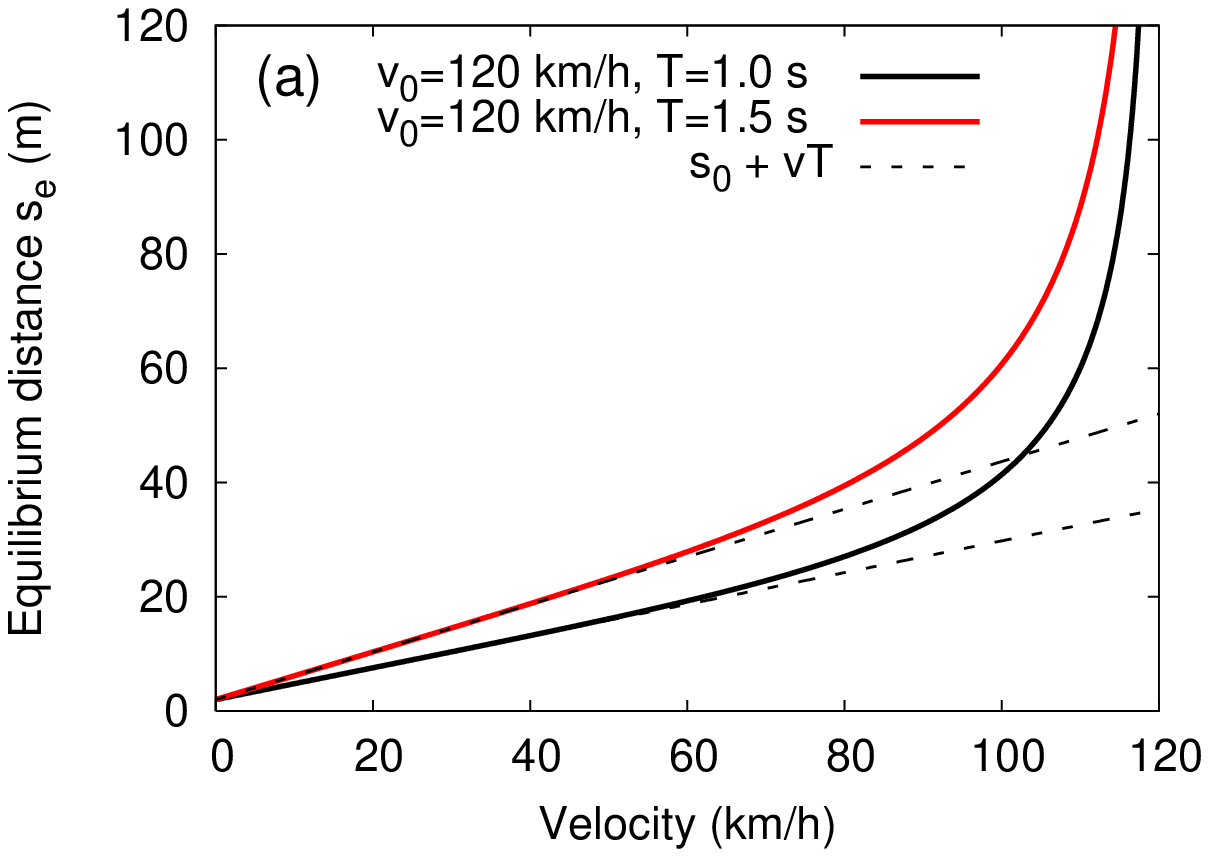}}
\includegraphics[width=60mm]{\cpath{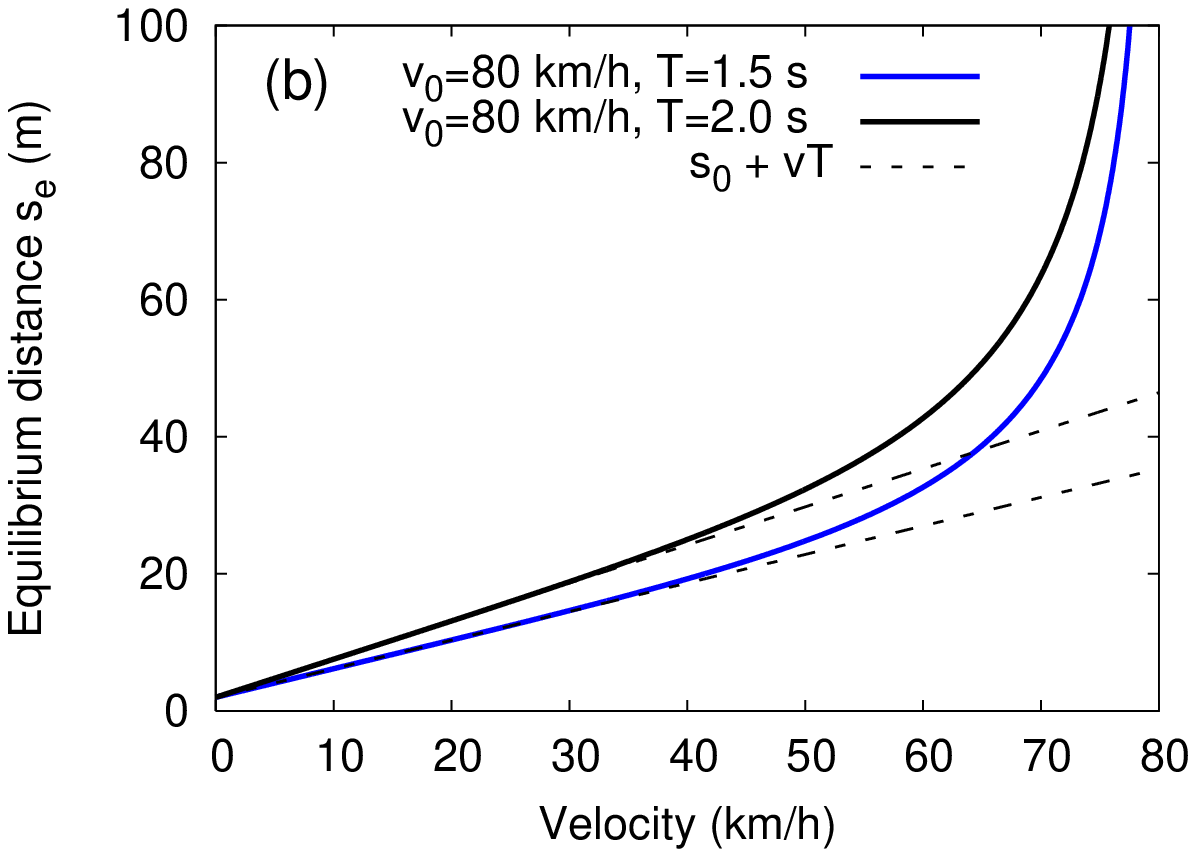}}

 \caption{\label{fig:IDM_equil}Equilibrium distance $s_e(v)$ according
 to Eq.~\protect\eqref{eq:equil_idm} as functions of the speed for
 different settings of the desired speed $v_0$ and the safety time gap
 $T$. The deviations from the dotted lines are discussed in the main
 text. The other parameters are those listed in the caption of
 Fig.~\ref{fig:IDM_single}.}

\end{figure}

The \textit{equilibrium properties} of the IDM are influenced by the
parameters for the desired time gap $T$ the desired speed $v_0$ and
the minimum distance between vehicles at a standstill
$s_0$. Equilibrium traffic is defined by vanishing speed differences
and accelerations of the driver-vehicle agents $\alpha$:
\begin{eqnarray}\label{eq:equi_conditions}
\Delta v_\alpha &=& 0,\\
\ableitung{v_\alpha}{t}  &=& 0,\\
\text{and} \quad \ableitung{v_{\alpha-1}}{t}  &=& 0.
\end{eqnarray}
Under these stationary traffic conditions, drivers tend to keep a
velocity-dependent {\it equilibrium gap} $s_e(v_\alpha)$ to the
leading vehicle. In the following, we consider a homogeneous ensemble
of identical driver-vehicle agents corresponding to identical parameter
settings. Then, the IDM acceleration equation~\eqref{eq:IDMaccel} with
the constant setting $\delta=4$ simplifies to
\begin{equation}\label{eq:equil_idm}
s_{e}(v) = \frac{s_0+vT}{\sqrt{1 - \left(\frac{v}{v_0}\right)^{4}}}.
\end{equation}
The equilibrium distance depends only on the minimum jam distance
$s_0$, the safety time gap $T$ and the desired speed $v_0$. The
diagrams (a) and (b) in Fig.~\ref{fig:IDM_equil} show the equilibrium
distance as a function of the velocity, $s_e(v)$, for different $v_0$
and $T$ parameter settings while keeping the minimum distance constant
at $s_0=2\,$m. In particular, the equilibrium gap of homogeneous {\it
congested} traffic ({with} $v \ll v_0$) is essentially equal to the
desired gap, $s_{e}(v) \approx s^*(v,0)=s_0+v T$. It is therefore
composed of the minimum bumper-to-bumper distance $s_0$ kept in
stationary traffic at $v=0$ and an additional velocity-dependent
contribution $v T$ corresponding to a constant safety time gap $T$ as
shown in the diagrams by straight lines. For $v\to 0$, the equilibrium
distance approaches the minimum distance $s_0$. If the velocity is
close to the desired speed, $v\approx v_0$, the equilibrium distance
$s_e$ is clearly larger than the distance $v T$ according to the
safety time gap parameter. For $v\to v_0$, the equilibrium distance
diverges due to the vanishing denominator in
Eq.~\eqref{eq:equil_idm}. That is, the free speed is reached {\it
exactly} only on a free road.

\begin{figure}[t]
\centering
\includegraphics[width=60mm]{\cpath{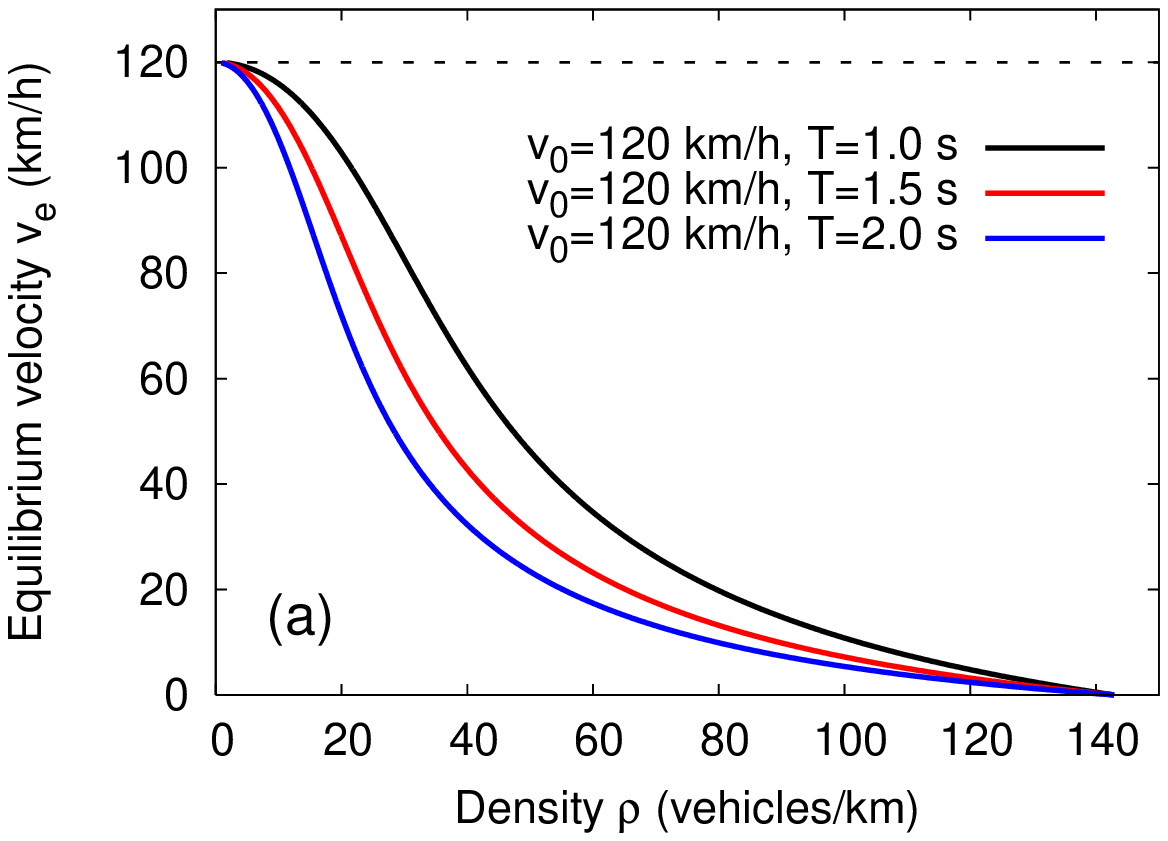}} 
\includegraphics[width=60mm]{\cpath{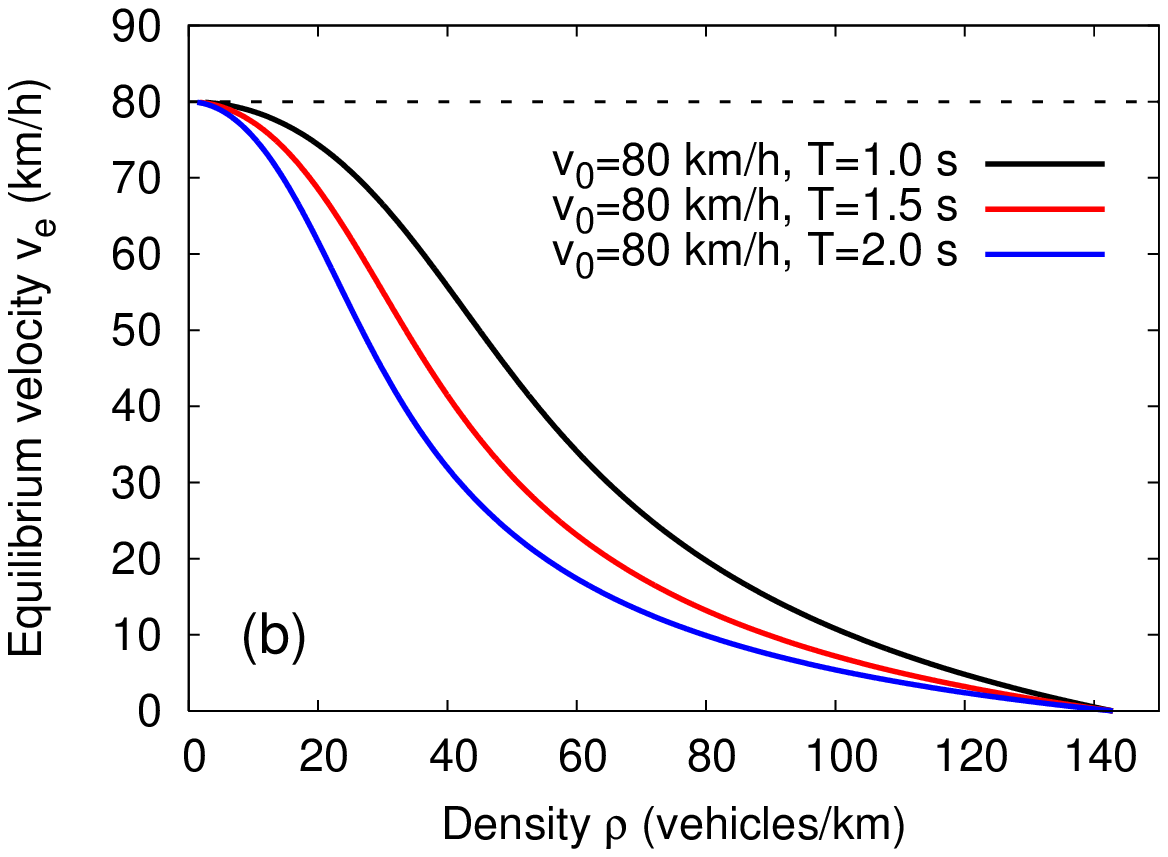}}\\
\includegraphics[width=60mm]{\cpath{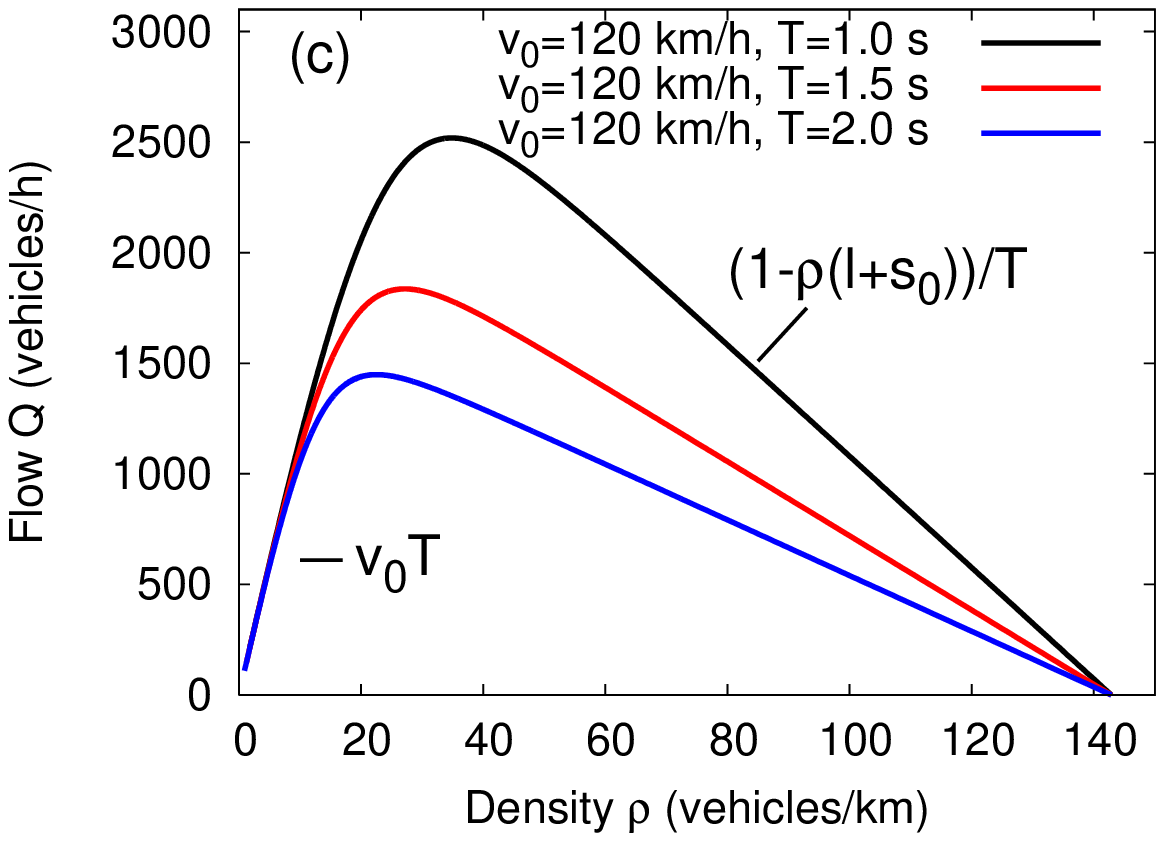}} 
\includegraphics[width=60mm]{\cpath{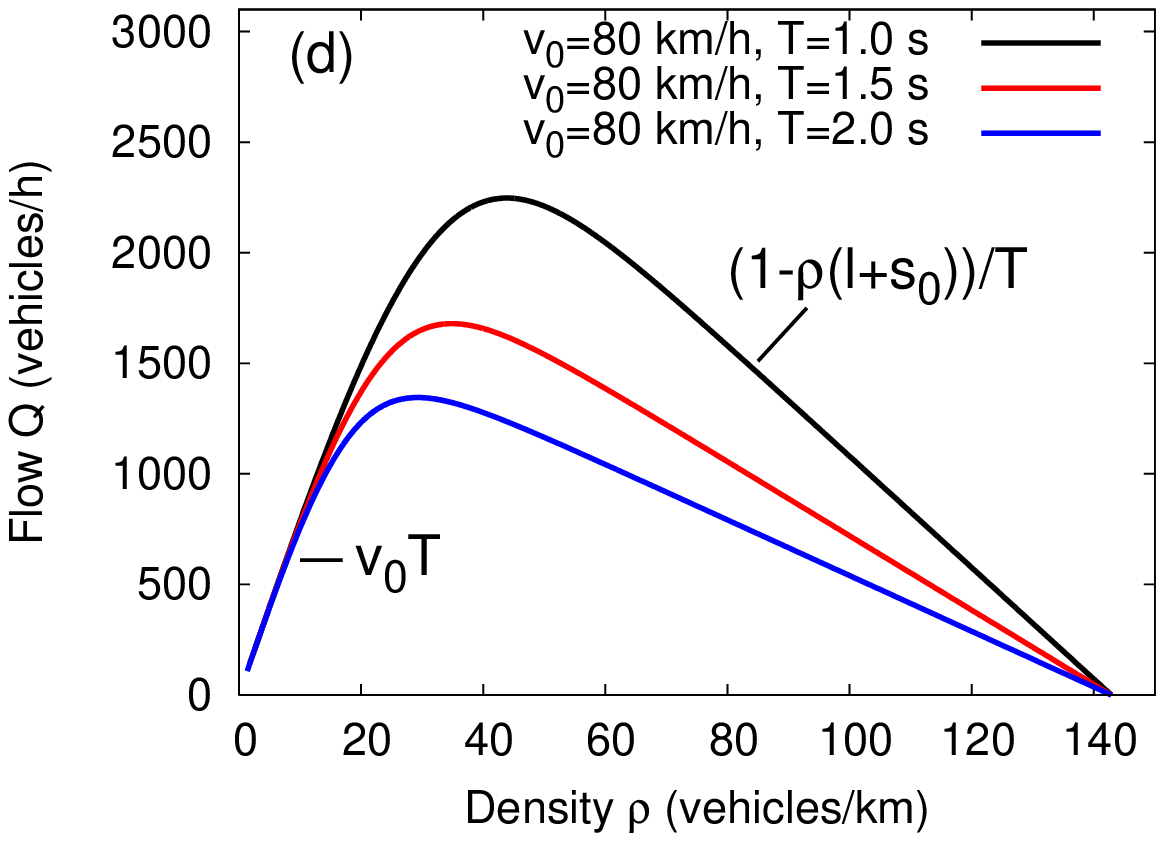}}

 \caption{\label{fig:IDM_fd}Equilibrium velocity-density relations of
 the IDM (top) and corresponding flow-density relations, so-called
 {\it fundamental diagrams} (bottom). The equilibrium properties
 depend on the minimum distance $s_0$ (here set to \unit[2]{m}), the
 desired speed $v_0$ (here displayed for 120 and \unit[80]{km/h}) and
 the time gap $T$ (here \unit[1.0, 1.5 and 2.0]{s}). The safety time gap
 is the most important parameter determining the maximum flow
 (stationary freeway capacity).}

\end{figure}

In the literature, the equilibrium state of homogeneous and stationary
traffic is often formulated in macroscopic quantities such as traffic
flow $Q$, (local) average velocity $V$ and traffic density
$\rho$. The translation from the microscopic net distance $s$ into the
density is given by the {\it micro-macro relation}
\begin{equation}\label{eq:srho}
s = \frac{1}{\rho} - l,
\end{equation} 
where $l$ is the vehicle length. In equilibrium traffic, $\rho$ is
therefore given by $s_e$, the mean velocity is simply $V=v_e$ and the
traffic flow follows from the hydrodynamic relation
\begin{equation}\label{eq:hydrodyn}
Q =\rho\, V.
\end{equation}
So, the equilibrium velocity $v_e$ is needed as a function of the
distance $s_e$. An analytical expression for the inverse of
Eq.~\eqref{eq:equil_idm}, that is the {\it equilibrium velocity} as a
function of the gap, $v_e(s)$, is only available for the acceleration
exponents $\delta=1,2$ or $\delta\to \infty$~\cite{Opus}. For
$\delta=4$, we only have a parametric representation $\rho(v)$ with
$v\in [0, v_0]$ resulting from Eqs.~\eqref{eq:srho} and~\eqref{eq:equil_idm}.
Figures~\ref{fig:IDM_fd}(a) and (b) show the equilibrium
velocity-density relation $V_e(\rho)$ for the same parameter settings
as in Fig.~\ref{fig:IDM_equil}. The assumed vehicle length $l=5\,$m
together with the minimum jam distance $s_0=2\,$m results in a
maximum density $\rho_\text{max}=1/(s_0+l)\approx
143\,$vehicles/km. Using the relation~\eqref{eq:hydrodyn}, we obtain
the so-called {\it fundamental diagram} between the traffic flow and
the vehicle density, $Q(\rho)=V \rho(v)$ which is displayed in Fig.~\ref{fig:IDM_fd}(c) and (d). Notice that
$Q$ is typically given in units of vehicles per hour and the density
$\rho$ in units of vehicles per km.

According to Eqs.~\eqref{eq:equil_idm} and~\eqref{eq:srho}, the {\it
fundamental relations} of homogeneous traffic depend on the desired
speed $v_0$ (low density), the safety time gap $T$ (high density) and
the jam distance $s_0$ (jammed traffic). In the low-density limit
$\rho\ll 1/(v_0 T)$, the equilibrium flow can be approximated by
$Q\approx v_0\rho$. In the high density regime, one has a linear
decrease of the flow,
\begin{equation}\label{eq:q_cong}
Q(\rho) \approx \frac{1-\rho(l+s_0)}{T},
\end{equation}
which can be used to determine the effective length $l+s_0$ and
$T$. Notice that the vehicle length is not a model parameter but only
a scaling quantity that determines the (static) maximum density
$\rho_\text{max}$ together with the IDM parameter $s_0$.

\subsection{Inter-Driver Variability}\label{sec:inter}

An important aspect of vehicular traffic is the heterogeneity of
agents, a term which includes characteristics of the drivers as well
as features of the vehicle. The multi-agent simulation approach is
appropriate for representing this heterogeneity on a microscopic
level. In order to address \textit{inter-driver variability}
(different drivers behave differently in identical traffic situations)
and vehicle properties (such as length, width, weight and
motorization) we propose to group driver-vehicle agents into classes
defining their specific driving styles and vehicle properties. For
this purpose, it is advantageous that the parameters of the
Intelligent Driver Model do have an intuitive meaning and are directly
related to driving behavior. In the following, we discuss the
parameter settings for three classes of passenger car drivers
representing ``normal'', ``timid'' and ``aggressive'' driving
styles. In addition, we model a typical truck driver. The corresponding
parameter values are listed in Table~\ref{tab:IDM}.

\begin{itemize}
\item
The \textit{desired speed} $v_0$ is the maximum speed a driver-vehicle
agent aims to reach under unobstructed driving conditions. A natural
value and upper limit for this parameter would be the typical
(highest) speed on the considered road element. The normal driver
chooses for instance~\unit[120]{km/h} on a freeway while a timid
driver prefers a lower value and a more aggressive driver likes to go
faster. The desired speed could be limited by legislation. In city
traffic, the speed is typically limited to \unit[50]{km/h} (cf.\ the
simulation scenario in Sec.~\ref{sec:trafficlight}). In this case, a
timid driver likes to drive a bit below this limit while an aggressive
driver can easily be modeled by an individual ``disobedience
factor''. Notice that strict speed limits apply to trucks on the whole
road network in most countries.
\item
The \textit{desired time gap} $T$ refers to the preferred distance $v
T$ while driving at speed $v$, cf.\ Eq.~\eqref{eq:IDMsstar}, and
mainly determines the maximum capacity (cf.\ Fig.~\ref{fig:IDM_fd}).
A typical value in dense traffic is about
\unit[1.4]{s} while German road authorities recommend~\unit[1.8]{s}. A
common observation on European freeways is that very small time gaps
are kept~\cite{VDT,Kno02-data}.
\item
The parameter $s_0$ describes the \textit{minimum bumper-to-bumper
distance} at a standstill, cf.\ Eq.~\eqref{eq:IDMsstar}. Typical gaps
in a queue of vehicles standing at traffic lights are in the range
between \unit[1]{m} and \unit[5]{m}. While a normal driver typically
keeps a minimum gap of \unit[2]{m}, a cautious driver prefers larger
gaps and an aggressive driver likes tailgating. It is natural to
assume that truck drivers prefer slightly larger gap than the normal car
driver due to larger vehicle dimensions. Notice that the vehicle
length is not a model parameter. However, it determines the maximum
density together with the minimum distance $s_0$ according to
Eq.~\eqref{eq:srho}. Typical vehicle lengths are for
instance~\unit[5]{m} for cars and \unit[12]{m} for trucks.
\item
The \textit{desired acceleration} $a$ describes the acceleration
behavior of the driver. Notice that the acceleration depends on the
actual vehicle speed as shown, for example, in
Fig.~\ref{fig:IDM_single}.  Since the acceleration \textit{behavior}
is based on a physical movement, the value of $a$ has to respect the
limits of motorization. Consequently, a truck has to be modeled by a
lower desired acceleration $a$ than a passenger car. An aggressive
driver prefers to accelerate fast (e.g.\ $\unit[3]{m/s^2}$) while a
timid driver prefers a lower value (e.g.\ $\unit[1]{m/s^2}$). The
acceleration exponent $\delta=4$ is kept constant for all driver
classes, cf.\ Eq.~\eqref{eq:IDMaccel}.
\item
The \textit{comfortable braking deceleration} $b$ determines the
approaching process toward slower leaders or stationary objects such
as traffic lights (see Sec.~\ref{sec:trafficlight}). As the IDM tries
to limit the braking deceleration to $b$, a low value
($b=\unit[1]{m/s^2}$) represents a driver who breaks accurately 
in an anticipative way corresponding to a smooth driving style. By way
of contrast, a higher value ($b=\unit[3]{m/s^2}$) describes an
aggressive driver who prefers to approach the leader with a large
velocity difference.
\end{itemize}
Taking these average parameters for each driver class as a starting
point, it is straightforward to distribute individual agent parameters
randomly within given limits, e.g.\ according to a uniform
distribution with a variation of 20\%.

\begin{table}
\centering
{\small
\begin{tabular}{lllll}
\toprule
IDM Parameter                       & Normal & Timid & Aggressive & Truck \\
\midrule
Desired speed $v_0$ in km/h         & 120 & 100 & 140 & 85 \\
Desired time gap $T$ in s           & 1.5 & 1.8 & 1.0 & 2.0 \\
Jam distance $s_0$ in m             & 2.0 & 4.0 & 1.0 & 4.0 \\
Maximum acceleration $a$ in m/s$^2$ & 1.4 & 1.0 & 2.0 & 0.7 \\
Desired deceleration $b$ in m/s$^2$ & 2.0 & 1.0 & 3.0 & 2.0 \\
\bottomrule
\end{tabular}
}
 \caption{\label{tab:IDM}Model parameters of the Intelligent Driver
 Model for three classes of passenger car drivers and a typical truck
 driver.}

\end{table}

\subsection{Intra-Driver Variability}\label{sec:memory}

Besides reacting to the immediate traffic environment, human drivers
adapt their driving style on longer time scales to the traffic
situation. Thus, the actual driving style depends on the traffic
conditions of the last few minutes which we call \textit{memory
effect}~\cite{IDMM}. For example, it is observed that most drivers
increase their preferred temporal headway after being stuck in
congested traffic for some time~\cite{Brilon-traff95,TGF01}.
Furthermore, when larger gaps appear or when reaching the downstream
front of the congested zone, human drivers accelerate less and
possibly decrease their desired speed as compared to a free-traffic
situation. 

In contrast to inter-driver variability considered in
Sec.~\ref{sec:inter}, the memory effect is an example of 
\textit{intra-driver variability} meaning that a driver behaves
differently in similar traffic situations depending on his or her
individual driving history and experience. Again, the multi-agent
approach can easily cope with this extension of driving behavior as
soon as one has a specific model to implement. By way of example, we
present a model that introduces a time-dependency of some parameters
of the Intelligent Driver Model to describe the \textit{frustration}
of drivers being in a traffic jam for a period~\cite{IDMM}.

We assume that adaptations of the driving style are controlled by a
single internal dynamical variable $\lambda(t)$ that represents the
``subjective level of service'' ranging from 0 (in a standstill)
to~1 (on a free road). The subjective level of service $\lambda(t)$
relaxes to the instantaneous level of service $\lambda_0(v)$ depending
on the agent's speed $v(t)$ with a relaxation time $\tau$ according to
\begin{equation}
\frac{\text{d}\lambda}{\text{d}t} = \frac{\lambda_0(v) - \lambda}{\tau}.
\end{equation}
This means that for each driver, the subjective level of service is given
by the exponential moving average of the instantaneous level of
service experienced in the past: 
\begin{equation}
\lambda(t) = \int_0^t \, \lambda_0(v(t'))\, e^{-(t-t')/\tau} \text{d}t'.
\end{equation}
We have assumed the instantaneous level of service $\lambda_0(v)$ to
be a function of the actual velocity $v(t)$. Obviously, $\lambda_0(v)$
should be a monotonically increasing function with $\lambda_0(0)=0$
and $\lambda_0(v_0)=1$ when driving with the desired speed $v_0$. The
most simple ``level-of-service function'' satisfying these conditions
is the linear relation
\begin{equation}
\lambda_0(v) = \frac{v}{v_0}.
\end{equation}
Notice that this equation reflects the level of service or efficiency
of movement from the agent's point of view, with $\lambda_0=1$ meaning
zero hindrance and $\lambda_0=0$ meaning maximum hindrance.  If one
models inter-driver variability (Sec.~\ref{sec:inter}) where different
drivers have different desired velocities, there is no objective level
of service, but rather only an individual and an average one.

Having defined how the traffic environment influences the degree of
adaptation $\lambda(t)$ of each agent, we now specify how this
internal variable influences driving behavior. A behavioral
variable that is both measurable and strongly influences the traffic
dynamics is the desired time gap $T$ of the IDM. It is observed that,
in congested traffic, the whole distribution of time gaps is shifted
to the right when compared with the data of free traffic~\cite{IDMM,VDT}. We
model this increase by varying the corresponding IDM parameter in the
range between $T_0$ (free traffic) and $T_\text{jam}=\beta_T T_0$
(traffic jam) according to
\begin{equation}
T(\lambda) = \lambda T_0 + (1-\lambda) T_\text{jam} = T_0 \left[
\beta_T + \lambda(1-\beta_T) \right].
\end{equation}
Herein, the adaptation factor $\beta_T$ is a model parameter. A value
for the frustration effect is $\beta_T=T_\text{jam}/T_0 = 1.8$ which
is consistent with empirical observations. A typical relaxation time
for the driver's adaptation is $\tau=\unit[5]{min}$.
Notice that other parameters of the driving style are probably
influenced as well, such as the acceleration $a$, the comfortable
deceleration $b$ or the desired velocity $v_0$. This could be
implemented by analogous equations for these parameters. Furthermore,
other adaption processes as well as the presented frustration effect
are also relevant~\cite{VDT}.

\section{Modeling Discrete Decisions}\label{sec:discrete}

On the road network, drivers encounter many situations where a
decision between two or more alternatives is required. This relates
not only to lane-changing decisions but also to considerations as to
whether or not it is safe to enter the priority road at an
unsignalized junction, to cross such a junction or to start an
overtaking maneuver on a rural road. Another question concerns whether
or not to stop at an amber-phase traffic light.  All of the above
problems belong to the class of \textit{discrete-choice problems}
that, since the pioneering work of McFadden~\cite{Hausman-MNL}, has
been extensively investigated in an economic context as well as in the
context of transportation planning. In spite of the relevance to
everyday driving situations, there are fewer investigations attempting
to incorporate the aforementioned discrete-choice tasks into
microscopic models of traffic flow, and most of them are restricted to
modeling lane changes~\cite{Gipps86}.  Only very recently acceleration
and discrete-choice tasks have been treated more
systematically~\cite{toledo2007,MOBIL-TRR07}.

The modeling of lane changes is typically considered as a multi-step
process. On a \textit{strategic} level, the driver knows about his or
her route on the network which influences the lane choice, e.g.\ with
regard to lane blockages, on-ramps, off-ramps or other mandatory
merges~\cite{Toledo-TRR-05}. In the \textit{tactical} stage, an
intended lane change is prepared and initiated by advance
accelerations or decelerations of the driver, and possibly by
cooperation of drivers in the target lane~\cite{Hidas-05}. Finally, in
the \textit{operational} stage, one determines if an immediate lane
change is both safe and desired~\cite{Gipps86}. While mandatory
changes are performed for strategic reasons, the driver's motivation
for discretionary lane changes is a perceived improvement of the
driving conditions in the target lane compared with the current 
situation.

In the following, we will present a recently formulated general
framework for modeling traffic-related discrete-choice situations in
terms of the acceleration function of a longitudinal
model~\cite{MOBIL-TRR07}.  For the purpose of illustration, we will
apply the concept to model mandatory and discretionary lane changes
(Sec.~\ref{sec:MOBIL}). Furthermore, we will consider the decision
process whether or not to brake  when approaching a traffic light
turning from green to amber (Sec.~\ref{sec:trafficlight}).

\subsection{Modeling Lane Changes}\label{sec:MOBIL}

Complementary to the longitudinal movement, lane-changing is a
required ingredient for simulations of multi-lane traffic. The
realistic description of multi-agent systems is only possible within a
multi-lane modeling framework allowing faster driver-vehicle agents to
improve their driving conditions by passing slower vehicles.
 
When considering a lane change, a driver typically makes a trade-off
between the expected own advantage and the disadvantage imposed on
other drivers. For a driver considering a lane change, the subjective
utility of a change increases with the gap to the new leader in the
target lane. However, if the speed of this leader is lower, it may be
favorable to stay in the present lane despite the smaller gap. A
criterion for the utility including {\it both} situations is the
difference between the accelerations after and before the lane change.
This is the core idea of the lane-changing algorithm
MOBIL~\cite{MOBIL-TRR07} that is based on the expected (dis)advantage
in the new lane in terms of the difference in the acceleration which
is calculated with an underlying microscopic longitudinal traffic
model, e.g.\ the Intelligent Driver Model (Sec.~\ref{sec:IDM}).

For the lane-changing decision, we first consider a \textit{safety
constraint}. In order to avoid accidents by the follower in the
prospective target lane, the safety criterion
\begin{equation}\label{eq:safety} 
\dot{v}_\text{follow} \ge -b_\text{safe}
\end{equation} 
guarantees that the deceleration of the successor
$\dot{v}_\text{follow}$ in the target lane does not exceed a safe
limit $b_\text{safe} \simeq 4\,$m/s$^2$ after the lane change. In
other words, the safety criterion essentially restricts the
deceleration of the lag vehicle on the target lane to values below
$b_\text{safe}$. Although formulated as a simple inequality, this
condition contains all the information provided by the longitudinal
model via the acceleration $\dot{v}_\text{follow}$. In particular, if
the longitudinal model has a built-in sensitivity with respect to
\textit{velocity differences} (such as the IDM) this dependence is
transfered to the lane-changing decisions. In this way, larger gaps
between the following vehicle in the target lane and the own position
are required to satisfy the safety constraint if the speed of the
following vehicle is higher than the own speed. In contrast, lower
values for the gap are allowed if the back vehicle is
slower. Moreover, by formulating the criterion in terms of safe
braking decelerations of the longitudinal model, crashes due to lane
changes are \textit{automatically} excluded as long as the
longitudinal model itself guarantees crash-free dynamics.

For discretionary lane changes, an additional \textit{incentive
criterion} favors lane changes whenever the acceleration in one of the
target lanes is higher. The incentive criterion for a lane change is
also formulated in terms of accelerations. A lane change is executed
if the sum of the own acceleration and those of the affected
neighboring vehicle-driver agent is higher in the prospective
situation than in the current local traffic state (and if the safety
criterion~\eqref{eq:safety} is satisfied of course). The innovation of
the MOBIL framework~\cite{MOBIL-TRR07} is that the immediately
affected neighbors are considered by the ``politeness factor''
$p$. For an egoistic driver corresponding to $p=0$, this incentive
criterion simplifies to $\dot{v}_\text{new} >
\dot{v}_\text{old}$. However, for $p=1$, lane changes are only carried
out if this increases the combined accelerations of the lane-changing
driver and all affected neighbors. This strategy can be paraphrased by
the acronym \textit{``Minimizing Overall Braking Induced by Lane
Changes''} (MOBIL). We observed realistic lane-changing behavior for
politeness parameters in the range $0.2 < p <
1$~\cite{MOBIL-TRR07}. Additional restrictions can easily be
included. For example, the ``keep-right'' directive of most European
countries is implemented by adding a bias to the incentive
criterion. A ``keep-lane'' behavior is modeled by an additional
constant threshold when considering a lane change.

\subsection{Approaching a Traffic Light}\label{sec:trafficlight}
\begin{figure}[t]
\centering
\includegraphics[width=70mm]{\cpath{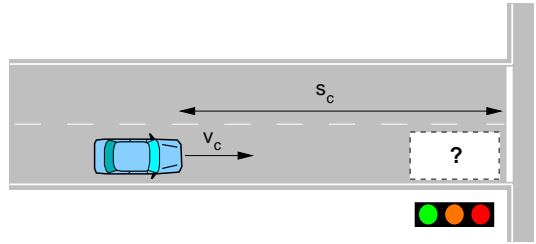}}

  \caption{\label{fig:traffLight}Approaching a traffic switching from
  green to amber. The two options of the decision situation are to
  stop in front of the light or to pass the amber-phase traffic light with
  unchanged speed.}

\end{figure}

When approaching a traffic light that switches from green to amber, a
decision has to be made whether to stop just at the traffic light or
to pass the amber-phase light with unchanged speed. For an empirical
study on the stopping/running decision at the onset of an amber phase
we refer to Ref.~\cite{rakha-trafficlights-2007}. If the first option
is selected, the traffic light will be modeled by a standing
``virtual'' vehicle at the position of the light. Otherwise, the
traffic light will be ignored. The criterion is satisfied for the
``stop at the light'' option if the own braking deceleration at the
time of the decision does not exceed the safe deceleration
$b_\text{safe}$. The situation is illustrated in
Fig.~\ref{fig:traffLight}. Denoting the distance to the traffic light
by $s_c$ and the velocity at decision time by $v_c$ and assuming a
longitudinal model of the form~\eqref{eq:IDMaccel}, the safety
criterion~\eqref{eq:safety} can be written as
\begin{equation}\label{eq:safety-traffLight}
\dot{v}(s_c,v_c,v_c) \ge -b_\text{safe}.
\end{equation}
Notice that the approaching rate and the velocity are equal ($\Delta
v_c=v_c$ ) in this case. The incentive criterion is governed by the
bias towards the stopping decision because legislation requires that
one stop at an amber-phase traffic light if it is safe to do
so. As a consequence, the incentive criterion is always fulfilled, and
Eq.~\eqref{eq:safety-traffLight} is the only decision criterion in
this situation.

Similarly to the lane-changing rules, the ``stopping
criterion''~\eqref{eq:safety-traffLight} will inherit all the
sophistication of the underlying car-following model. In particular,
when using realistic longitudinal models, one obtains a realistic
stopping criterion with only one additional parameter
$b_\text{safe}$. Conversely, unrealistic microscopic models such as
the Optimal Velocity Model~\cite{Bando} or the Nagel-Schreckenberg
cellular automaton~\cite{Nagel-S} will lead to unrealistic
stopping-decisions. In the case of the Optimal Velocity Model, it is
not even guaranteed that drivers deciding to stop will be able to stop
at the lights.

\begin{figure}[t]
\centering
\includegraphics[width=70mm]{\cpath{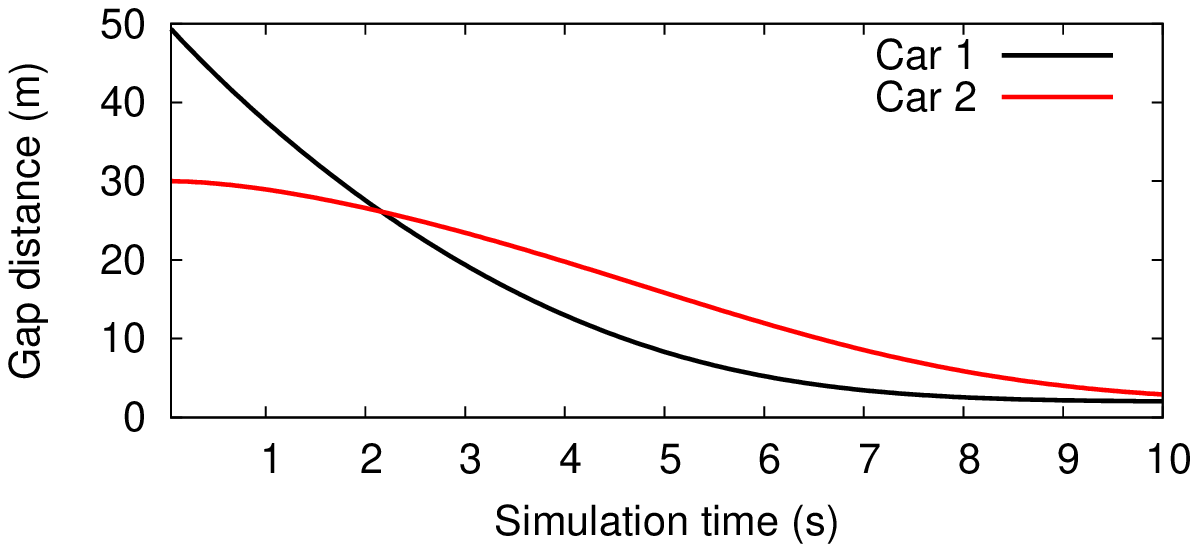}}\\
\includegraphics[width=70mm]{\cpath{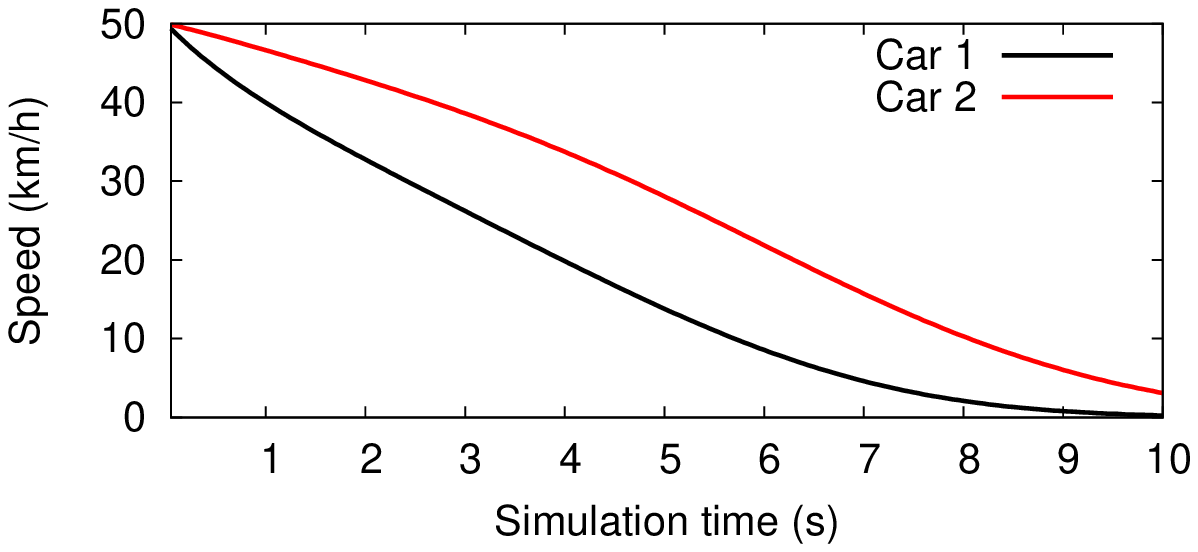}}\\
\includegraphics[width=70mm]{\cpath{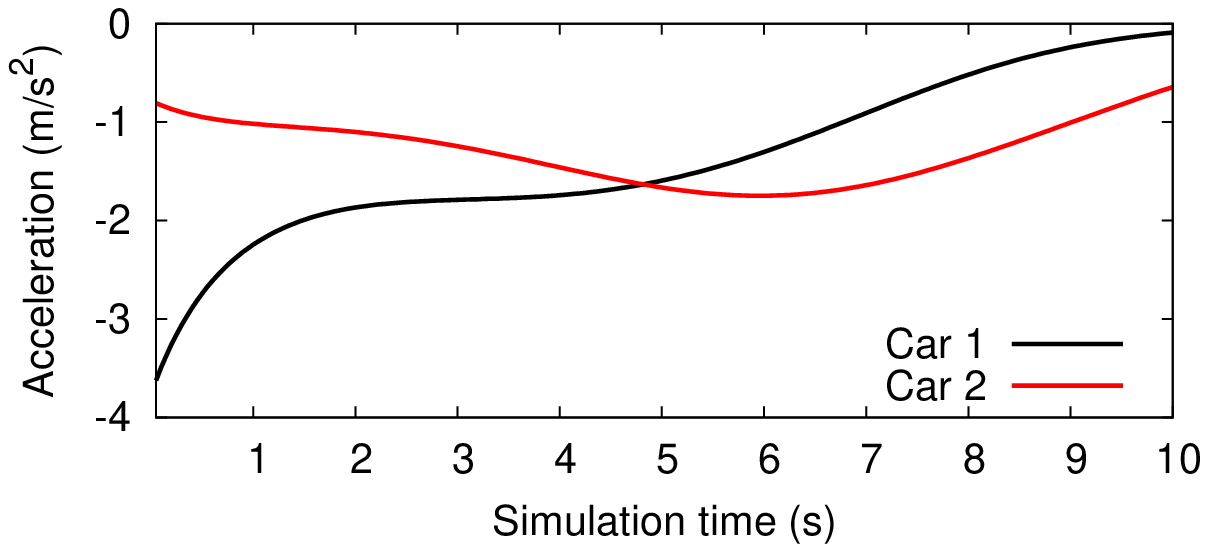}}

  \caption{\label{fig:IDMdecel}Maneuver of approaching a traffic light
  initially~\unit[50]{m} with a speed of~\unit[50]{km/h} according to
  the Intelligent Driver Model. The braking deceleration is limited to
  the comfortable braking deceleration (IDM parameter $b$) whenever
  possible. The stronger braking of the first car is needed to keep
  the situation under control. The parameters for the simulation are
  listed in Table~\ref{tab:IDM}.}

\end{figure}

For the purpose of illustration, we apply the concept to the following
situation in city traffic: A car is driving at speed
$v_c=\unit[50]{km/h}$ towards an amber traffic light located at a
distance $s_c=\unit[50]{m}$. Applying the IDM parameters of a
``normal'' driver listed in Table~\ref{tab:IDM} in combination with an
adapted desired speed of $v_0=\unit[50]{km/h}$, the acceleration
function~\eqref{eq:IDMaccel} results in an initial braking of
$\dot{v}(0)\approx \unit[3.6]{m/s^2}$ at $t=\unit[0]{s}$. For a safe
deceleration equal to the desired deceleration of the IDM, that is
$b_\text{safe}=b=\unit[2.0]{m/s^2}$, the MOBIL decision says ``drive
on''.  If, however, a safe braking deceleration of
$b_\text{safe}=\unit[4]{m/s^2}$ is assumed, the driver agent would
decide to brake resulting in the approaching maneuver shown in
Fig.~\ref{fig:IDMdecel}. The initial braking stronger than
$-\unit[2]{m/s^2}$ makes the situation manageable for the agent. After
\unit[2]{s}, the situation is ``under control'' and the vehicle brakes
approximately with the comfortable deceleration
$b=\unit[2]{m/s^2}$. In order to reach a standstill in a smooth way,
the deceleration is reduced to limit the \textit{jerk} which defines
the change in the acceleration. In addition, Fig.~\ref{fig:IDMdecel}
shows the behavior of the second vehicle following the leader. The
acceleration time series shows the important feature of the IDM in
limiting braking decelerations to the comfortable limit $b$ as long as
safety is warranted. From these results it is obvious that the setting
$b_\text{safe}=b$ is a natural assumption to model the decision
process realistically. Notice, however, that a human reaction time of
about \unit[1]{s}~\cite{green-reactionTimes} has to be taken into
account as well.

\section{Microscopic Traffic Simulation Software}\label{sec:simulator}

\begin{figure}[t]
\centering
\includegraphics[width=90mm]{\cpath{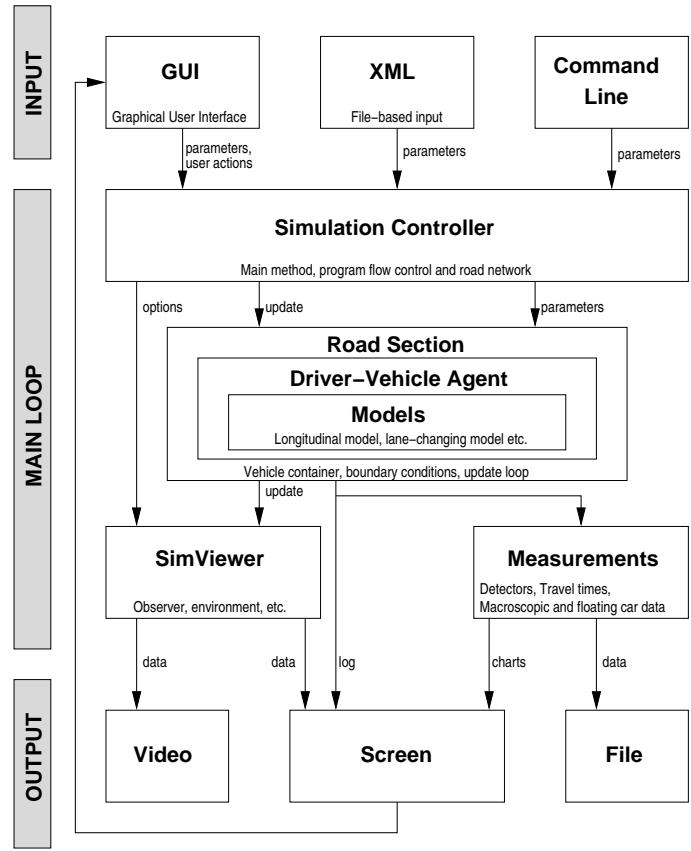}}

 \caption{\label{fig:SimOverview}Illustration of possible relations
 among functions in a traffic simulation framework. The input data
 defining a simulation setting can be provided by data files, the
 command line or a graphical user interface (GUI). The main simulation
 loop is organized by a ``Simulation Controller'' which controls the
 update of the road network, the graphical representation
 (``SimViewer'') and the output functions corresponding to
 measurements of several microscopic and macroscopic quantities.}

\end{figure}

So far, we have discussed models describing the longitudinal movement
and discrete decisions of individual driver-vehicle agents. Let us now
address the issue of a simulation framework that integrates these
components into a microscopic multi-lane traffic simulator. Typical
relations among functions in a microscopic traffic simulator are shown
in Fig.~\ref{fig:SimOverview}. On the level of input data, simulation
settings can be provided by input files, e.g.\ encoded in XML, by
command line or via a graphical user interface (GUI). The main
simulation loop is organized by a \textit{Simulation Controller} which
keeps track of the program operations and user actions. This central
control unit calls the update methods of the road-section objects. We
will elaborate on these components in Sec.~\ref{sec:design}. Since the
calculation of the vehicle accelerations is the very core of a traffic
simulation, we will pinpoint the issue in Sec.~\ref{sec:scheme}.
Simulation results can be written to data files and, in addition,
visualized by 2D and 3D computer graphics on the screen (see
Sec.~\ref{sec:visu}). Furthermore, we will extend the simulator in
order to simulate inter-vehicle communication (see Sec.~\ref{sec:IVC}
below).

There are a number of interactive simulators available publicly. The
website~\cite{traffic-simulation2007} deploys the Intelligent Driver
Model~\cite{Opus} introduced in Sec.~\ref{sec:IDM} for cars and trucks
together with the lane-changing algorithm
MOBIL~\cite{MOBIL-TRR07}. This demonstrator simulates typical
bottleneck scenarios such as on-ramps, lane-closings, uphill grades
and or traffic lights. Another open source simulator for whole traffic
networks is
\textsc{SUMO}~\cite{SUMO}.  The software
uses the Krauss model~\cite{KrauzzWagnerGaw-97}. Recently,
\textsc{FreeSim} has been made
available to the public~\cite{FREESIM}. Furthermore, commercial
traffic simulation software tools (for instance
\textsc{VISSIM}$^\text{\textsc{\tiny TM}}$,
\textsc{AIMSUN}$^\text{\textsc{\tiny TM}}$ or
\textsc{PARAMICS}$^\text{\textsc{\tiny TM}}$) offer a variety of additional
modules such as emission or pedestrian models and interfaces, e.g.\
for controlling simulation runs by remote and for implementing
additional features. These commercial products incorporate
sophisticated virtual environment 3D engines. Note, however, that the
underlying models are generally not well documented.

\subsection{Simulator Design}\label{sec:design}

Next to the functional view shown in Fig.~\ref{fig:SimOverview}, a
hierarchical view can be used to represent the dependencies and
inherited properties which makes use of the object-oriented programming
paradigm by representing and abstracting functional units as
classes. The best example is the representation of a driver-vehicle
agent as an abstract class with several possible designs for human
drivers, vehicles equipped with adaptive cruise
control~\cite{ACC-Arne-TRR07,Arne-ACC-TRC} or even driverless ones as
recently demonstrated in reality~\cite{darpa-07}.  However, each agent
has a number of defining properties such as length, width, weight,
form and color. Furthermore, each agent requires a model for the
lengthwise movement which is in turn an abstract class with the
presented IDM as a specific implementation. Further components are
required in order to model other aspects of driver behavior such as lane
changes, memory, etc. Since each agent is represented by an individual
object, it is straightforward to assign individual parameter values to
account for driver diversity (Sec.~\ref{sec:inter}).

The road network can be represented by connected road sections such as
main roads, on-ramps and off-ramps. A road section is defined by its
properties like length, number of lanes, etc. In addition, an element
may contain attributes representing the concrete infrastructure
relevant to the driver-vehicle agents such as lane closures, lane
narrowings, speed limits, uphill gradients and/or traffic lights. Notice
that the set of attributes which is relevant for the behavior and
decision-making has to be available to the agent.

The most detailed view on the innermost update loop of a road section
is given in terms of the following pseudo code:
\begin{verbatim}
updateRoadSection(){
  updateNeigborhood(); // organizing set of vehicles in multiple lanes   
  updateInfrastruture(); // active road attributes (e.g. traffic lights)
  updateAgentsRoadConditions(); // attributes affect agents
  calculateAccelerations(); //evaluate longitudinal models of agents 
  laneChanges();    // decision making and performing lane changes     
  updatePositionsAndSpeeds(); // integration within discrete update
  updateBoundaries(); // inflow and outflow
  updateOutput(); // log observable quantities and update detectors
}
\end{verbatim}

\subsection{Numerical Integration}\label{sec:scheme}
%
The explicit integration in the \texttt{updatePositionsAndSpeeds}
function of all driver-vehicle agents $\alpha$ is the very core of a
traffic simulator. In general, the longitudinal movement of the
vehicles is described by car-following models which take into account
the direct leader and result in expressions for the acceleration
function of the form
\begin{equation}\label{eq:App_mic}
\ableitung{v_{\alpha}}{t} = f\left(s_{\alpha}, v_{\alpha}, \Delta v_{\alpha}\right),
\end{equation}
that is the acceleration depends only on the own speed $v_{\alpha}$,
the gap $s_{\alpha}$, and the velocity difference (approaching rate)
$\Delta v_{\alpha}=v_{\alpha}-v_{\alpha-1}$ to the leading vehicle
$(\alpha-1)$. Note that we discussed the Intelligent Driver Model
(IDM) as an example for a car-following model in
Sec.~\ref{sec:IDM}. Together with the gap $s_{\alpha}(t)= x_{\alpha
-1}(t) - x_\alpha(t) - l_{\alpha-1}$ and the general equation of
motion,
\begin{equation}\label{motion}
\ableitung{x_{\alpha}}{t}=v_{\alpha},
\end{equation} 
Eq.~\eqref{eq:App_mic} represents a (locally) coupled system of
\textit{ordinary differential equations} (ODEs) for the positions $x_{\alpha}$ and
velocities $v_{\alpha}$ of all vehicles $\alpha$.  

As the considered acceleration functions $f$ are in general nonlinear,
we have to solve the set of ODEs by means of numerical integration. In
the context of car-following models, it is natural to use an explicit
scheme assuming constant \textit{accelerations} within each update
time interval $\Delta t$. This leads to the explicit numerical update
rules
\begin{equation}\label{eq:euler}
\begin{split}
v_{\alpha}(t+\Delta t) &= v_{\alpha}(t)+ \dot{v}_{\alpha}(t) \Delta t,\\
x_{\alpha}(t+\Delta t) &=
  x_{\alpha}(t)+ v_{\alpha}(t) \Delta t + \frac{1}{2} \dot{v}_{\alpha}(t) (\Delta t)^2,
\end{split}
\end{equation}
where $\dot{v}_{\alpha}(t)$ is an abbreviation for the acceleration
function $f\left(s_{\alpha}(t), v_{\alpha}(t), \Delta
v_{\alpha}(t)\right)$.  For $\Delta t \to \unit[0]{s}$, this scheme
locally converges to the exact solution of
\eqref{eq:App_mic} with consistency order 1 for the velocities (``Euler
update'', cf.\ \cite{NumRec}) and consistency order~2 for the
positions (``modified Euler update'') with respect to the
$L^2$-norm.\footnote{A time-continuous traffic model is mathematically
consistent if a unique local solution exists and if a numerical update
scheme exists whose solution locally converges to this solution when
the update time interval goes to zero. It has the consistency order
$q$ if $||\epsilon || = O(\Delta t^q)$ for $\Delta t\to \unit[0]{s}$
where $\epsilon$ denotes the deviation of the numerical solution for
$x_\alpha$ or $v_\alpha$ with respect to the exact solution, and
$||\cdot||$ is some functional norm such as the $L^2$-Norm.} Because
of the intuitive meaning of this update procedure in the context of
traffic, the update rule~\eqref{eq:euler} or similar rules are
sometimes considered to be part of the model itself rather than as a
numerical approximation~\cite{ThreeTimes-07}. A typical update time
interval $\Delta t$ for the IDM is between \unit[0.1]{s} and
\unit[0.2]{s}. Nevertheless, the IDM is approximately numerically
stable up to an update interval of $\Delta t \approx T/2$, that is half
of the desired time gap parameter $T$.


\subsection{Visualization}\label{sec:visu}
\begin{figure}
\centering
\includegraphics[width=0.7\linewidth]{\cpath{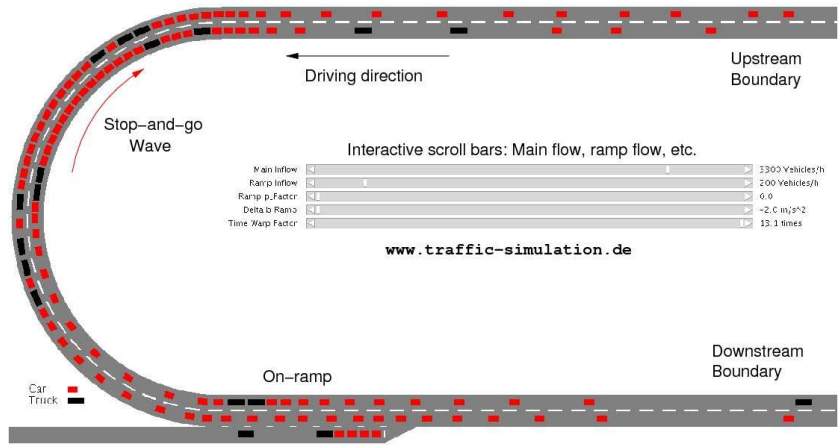}}

 \caption{\label{fig:2d_onramp}Example of 2D visualization of a
 two-lane freeway with an on-ramp from the
 website~\cite{traffic-simulation2007}. The screenshot shows the
 breakdown of traffic flow at the on-ramp serving as a bottleneck. A
 stop-and-go wave (cluster of vehicles) is propagating against the
 driving direction. The source code of the simulator is publicly
 available as an open source.}

\end{figure}

\begin{figure}
\centering
\includegraphics[width=85mm]{\cpath{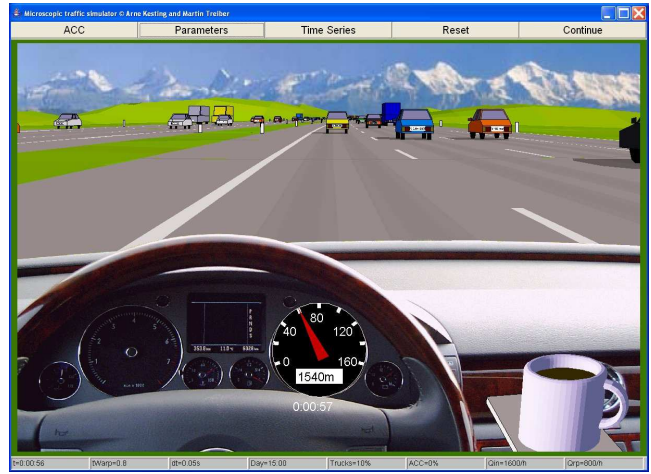}}

 \caption{\label{fig:3d_heli}Example of a 3D animation from the
 driver perspective. Notice that the ``Coffeemeter'' visualizes
 the acceleration and the jerk (the changes of all acceleration
 with time) which are difficult to visualize by other means. It
 therefore serves as a measure of the driving comfort.}

\end{figure}

%
Besides the implementation of the simulation controller with the focus
on quantitative models, the visualization of vehicle movements is also
an important aspect of simulation software.  In the case of vehicular
traffic it is straightforward to envision the vehicle trajectories over
the course of time whether in 2D or 3D. The latter representation is of course 
more demanding. Figure~\ref{fig:2d_onramp} illustrated an
example of a ``bird's eye view'' of a two-lane freeway with an
on-ramp, while Fig.~\ref{fig:3d_heli} illustrates a ``cockpit
perspective'' of a driving vehicle on the road. Note that the 3D
engine was programmed from scratch as an exercise by the
authors. However, higher level tools and open source 3D engines for
OpenGL are available. For more details on this subject we refer to the
chapter ``Crowd Behavior Modeling: From Cellular Automata to
Multi-Agent Systems'' by Bandini, Manzoni and Vizzari.

The animated visualization demonstrates both the individual
interactions and the resulting collective dynamics. In particular, the
graphical visualization turns out to be an important tool when
developing and testing lane-changing models and other decisions based
on complex interactions with neighboring vehicles for their
plausibility. In fact, the driving experiences of programmers offer
the best measure of realism and also provide stimulus for further
model improvements.

Last but not least, scientists and experts have to keep in mind that
computer animations have become an important tool for a fast and
intuitive knowledge transfer of traffic phenomena to students,
decision-makers and the public.  In particular, visualization in
\textit{real time} allows for direct user interaction influencing
the simulation run, e.g.\ by changing the simulation conditions (in
terms of inflows and driver population) and parameter settings of the
underlying models. In this way, the complexity of simulation
techniques (which are based on assumptions, mathematical models, many
parameters and implementational details) can become more accessible.

Finally, we remark that animation is a playground for the
programmers. For instance, the full cup of coffee in
Fig.~\ref{fig:3d_heli} represents not just a comforting habit of the
agent but is also a vivid way to illustrate the simulated longitudinal
acceleration as well as transverse accelerations due to
lane-changing. Moreover, the coffee level is a measure of the riding
comfort because it is also sensitive to the derivative of the
acceleration which is perceived as a jerk by the driver. The
hydrodynamic equations of the coffee surface in the cup with a
diameter $2r$ are astonishingly realistically approximated by a
harmonious pendulum with two degrees of freedom $\phi_x$ and $\phi_y$
denoting the angles of the surface normal:
\begin{equation}\label{eq_coffee}
\begin{split}
\ddot{\phi}_x + \frac{2\pi}{\tau}\dot{\phi}_x
+ \omega_0^2 \phi_x+\frac{\ddot{x}}{r} &= 0, \\
\ddot{\phi}_y + \frac{2\pi}{\tau}\dot{\phi}_y
+ \omega_0^2 \phi_y+\frac{\ddot{y}}{r} &= 0.
\end{split}
\end{equation}
The second time derivates $\ddot{x}$ and $\ddot{y}$ denote the vehicle
accelerations in longitudinal and transversal direction. The angular
frequency is $\omega_0 = \sqrt{g/r}$ where $g$ is the gravitational
constant.  During a coffee break, the damping time $\tau$ of Java
coffee was empirically determined as $\tau = \unit[12]{s}$ by the
authors.

\section{From Individual to Collective Properties}\label{sec:sim}

After having constructed the driver-vehicle agents, let us now adopt them
in a multi-agent simulation in which they interact with each
other. The process of simulating agents in parallel is one of
emergence from the microscopic level of pairwise interactions to the
higher, macroscopic level in order to reproduce and predict real
phenomena.

In this section, we will present three simulation applications. In
Sec.~\ref{sec:wave}, we will demonstrate the \textit{emergence of a
collective pattern from individual interactions} between
driver-vehicle agents by simulating the breakdown of traffic flow and
the development of a stop-and-go wave. The simulation will show the
expressive power of the Intelligent Driver Model in reproducing the
characteristic backwards propagation speed which is a well-known
constant of traffic world wide. In Sec.~\ref{sec:speedlimit}, we will
apply the traffic simulation framework to analyze the impact of a
speed limit as an example of a traffic control task . By way of this
example, we will demonstrate the predictive power of microscopic
traffic flow simulations. 

Last but not least, we apply the simulation framework to study a
coupled system consisting of communicating driver-vehicle agents using
short-range wireless networking technology in
Sec.~\ref{sec:IVC}. Since the multi-agent approach is a flexible
general-purpose tool, one can additionally equip an agent with
short-range communication devices that can self-organize with other
devices in range into ad-hoc networks. Such inter-vehicle
communication has recently gained much attention in the academic and
engineering world. It is expected to provide great enhancement to the
safety and efficiency of modern individual transportation systems. By
means of simulation, we will demonstrate the dissemination of
information about the local traffic situation over long distances even
for small equipment rates in the vehicle fleet.

\subsection{Emergence of Stop-and-Go Waves}\label{sec:wave}

\begin{figure}
\centering
\includegraphics[width=120mm]{\cpath{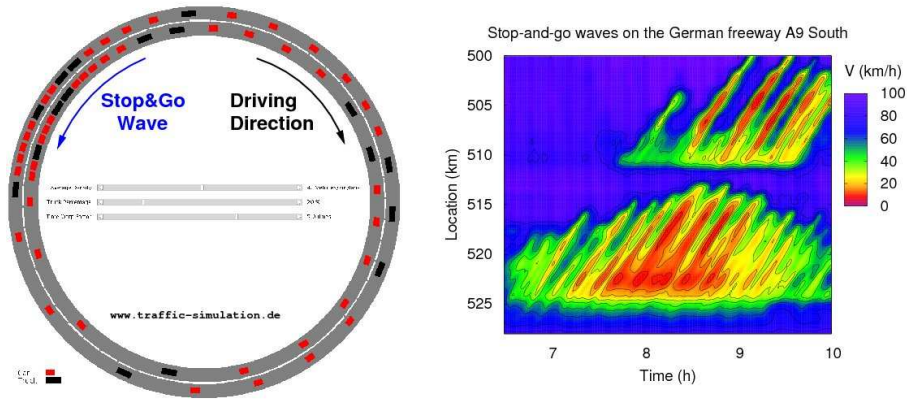}} 

 \caption{\label{fig:stopwave}(a) Emergence of a stop-and-go wave in a
 simulation and (b) stop-and-go traffic on the German freeway A9 South
 in the north Munich region. Notice that the spatiotemporal traffic
 dynamics in diagram (b) have been reconstructed from loop detector
 data using an interpolation method~\cite{Treiber-smooth}. The
 characteristic propagation speed of stop-and-go waves of about
 \unit[15]{km/h} against the driving direction is a self-organized
 ``constant'' of traffic flow which is reproduced by the Intelligent
 Driver Model used in the simulation (a).}

\end{figure}

Let us first study the emergence of a collective traffic phenomenon in
a simple ring road scenario as depicted in
Fig.~\ref{fig:stopwave}(a). Note that this scenario can be used 
interactively on the website~\cite{traffic-simulation2007}. Such
a closed system is defined by an initial value problem. The control
parameter is the homogeneous traffic density which essentially
determines the long-term behavior of the system. In the simulation,
the initial traffic density is too high to be able to retain free flow
conditions.

In the course of time, a vehicle eventually changes lanes resulting in
a smaller gap for the following vehicle, which, in turn, has to brake
in order to re-establish a safe distance to the new leader. After the
initial braking, the next follower again needs some time to respond to
this new situation by decelerating. The perturbation therefore
increases while propagating in upstream direction, that is against the
driving direction of the vehicle flow (see
Fig.~\ref{fig:stopwave}(a)). This response mechanism acts like a
``vicious circle'': Each following driver has to reduce his or her
speed a bit more to regain the necessary safety distance. Eventually,
vehicles further upstream in the platoon brake to a
standstill. Moreover, the time to re-accelerate to the restored speed
of the leading vehicle takes even more time due to limited
acceleration capabilities. Finally, we observe the emergence of a {\it
stop-and-go wave}.

Stop-and-go waves are also observed in real traffic as shown in
Fig.~\ref{fig:stopwave}(b) for the German freeway A9 South in the
north Munich region. Single stop-and-go waves propagate over more than
\unit[10]{km} leading to their description as ``phantom traffic
jams''. Their propagation speed is arguably constant. From the
time-space diagram in Fig.~\ref{fig:stopwave}(b), the propagation
speed of the downstream front of the stop-and-go wave can be
determined as approximately \unit[15.5]{km/h}. In each country,
typical values for this ``traffic constant'' are in the range
$\unit[15\pm 5]{km/h}$, depending on the accepted safe time clearance
and average vehicle length~\cite{Kerner-Rehb96-2}. Consequently,
realistic traffic models should reproduce this self-organized property
of traffic flow.

\subsection{Impact of a Speed Limit}\label{sec:speedlimit}

Microscopic traffic models are especially suited to the study of
heterogeneous traffic streams consisting of different and individual
types of driver-vehicle agents.  

\begin{figure}[t]
\centering

\includegraphics[width=0.8\linewidth]{\cpath{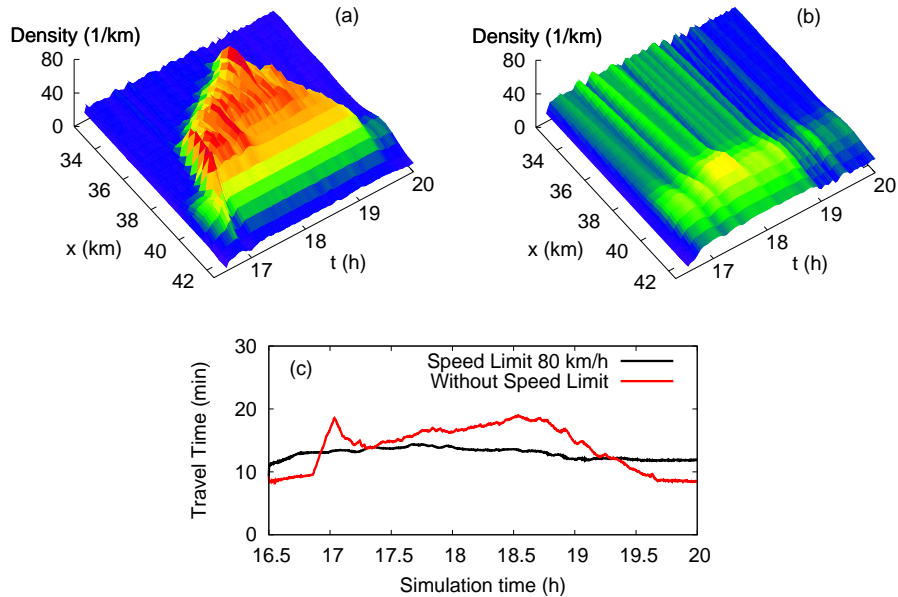}}

 \caption{\label{fig:speedl}Realistic simulation of an empirical
 traffic breakdown caused by an uphill gradient located around $x=$
 \unit[40]{km} by using velocity and flow data from loop detectors as
 upstream boundary conditions (a) without speed limit and (b) with a
 speed limit of \unit[80]{km/h}. Diagram (c) shows the travel times
 corresponding to the scenarios (a) and (b). Notice that the
 microscopic modeling approach allows for an estimation of the current
 travel time by simply summing up the travel times derived from the
 speeds of each driver-vehicle agent simultaneously. In accordance
 with Treiber and Helbing~\cite{Treiber-aut}.}
\end{figure}

In the following scenario, we will study the effect of a speed limit
for a section of the German freeway A8 East containing an uphill
section around $x=\unit[40]{km}$~\cite{Treiber-aut}. We considered the
situation during the evening rush hour on November 2, 1998. In the
evening rush hour at about \unit[17]{h}, traffic broke down at the
uphill section. In the simulation shown in Fig.~\ref{fig:speedl}(a),
we used lane-averaged one-minute data of velocity and flow measured by
loop detectors as upstream boundary conditions reproducing the
empirical traffic breakdown. In contrast to the ring road scenario in
Sec.~\ref{sec:wave}, the inflow at the upstream boundary is the natural
control parameter for the open system.

We have assumed two vehicle classes: 50\% of the drivers had a desired
speed of $v_0=\unit[120]{km/h}$, while the other half had
$v_0=\unit[160]{km/h}$ outside of the uphill region. A speed limit
reduces the desired velocities to \unit[80]{km/h}.  Within the uphill
region, both driver-vehicle classes are forced to drive at a maximum
of \unit[60]{km/h} (for example, due to overtaking trucks that are not
considered explicitly here).

Figure~\ref{fig:speedl} shows spatiotemporal plots of the locally
averaged traffic density for scenarios with and without the speed
limit. The simulations show the following: 
\begin{itemize}
 \item During the rush hour ($\unit[17]{h} \le t\le \unit[19]{h}$),
 the overall effect of the speed limit is positive. The increased
 travel times in regions without congestion are overcompensated by the
 saved time due to the avoided breakdown.  

 \item For lighter traffic ($t<\unit[17]{h}$ or $t>\unit[19:30]{h}$),
 however, the effect of the speed limit is clearly negative. Note that
 this problem can be circumvented by traffic-dependent, variable speed
 limits.
\end{itemize}

Although the speed limit reduces the velocity, it can improve the
quality of traffic flow while uphill regions obviously results in a
deterioration. To understand this counter-intuitive result, we point
out that the desired speed $v_0$ corresponds to the lowest value of
(i) the maximum velocity allowed by the motorization, (ii) the imposed
speed limits (possibly with a ``disobedience factor''), and (iii) the
velocity actually ``desired'' by the driver. Therefore, speed limits
act selectively on the \textit{faster} vehicles, while uphill
gradients reduce the speed especially of the
\textit{slower} vehicles. As a consequence, speed limits reduce
velocity differences, thereby stabilizing traffic, while uphill
gradients increase them.  For traffic consisting of
\textit{identical} driver-vehicle combinations (one driver-vehicle
class), these differences are neglected and both speed limits and
uphill gradients have in fact the same (negative) effect. Since global
speed limits always raise the travel time in off-peak hours when free
traffic is unconditionally stable (cf.\ Fig.~\ref{fig:speedl}(c)),
traffic-dependent speed limits are an optimal solution. Note that the
impact of a speed limit on the homogenization of traffic flow can be
studied interactively on the website~\cite{traffic-simulation2007} for
a lane-closing scenario instead of an uphill bottleneck.

\subsection{Store-and-Forward Strategy for Inter-Vehicle Communication}\label{sec:IVC}
%
Recently, there has been growing interest in wireless communication
between vehicles and potential applications. In particular,
inter-vehicle communication (IVC) is widely regarded as a promising
concept for the dissemination of information on the local traffic
situation and short-term travel time estimates for advanced traveler
information
systems~\cite{Jin-IVC-2006,IVC-TRC06,IVC-Goldrain06,IVC-Martin-TRR07,wisch}. In
contrast to conventional communication channels which operate with a
centralized broadcasting concept via radio or mobile phone services,
IVC is designed as a \textit{local service} based on the Dedicated
Short Range Communication standard enabling data transmission at a
frequency of \unit[5.8]{GHz}. These devices broadcast messages which
are received by all other equipped vehicles within a \textit{limited
broadcasting range}. As IVC message dissemination is not controlled by
a central station, no further communication infrastructure is
needed. For example, wireless local-area networks (IEEE 802.11 a/b/g)
have already shown their suitability for IVC with typical broadcasting
ranges of~\unit[200-500]{m}~\cite{Singh2002,OttKut2004}.

\begin{figure}
\centering
  \includegraphics[width=0.9\linewidth]{\cpath{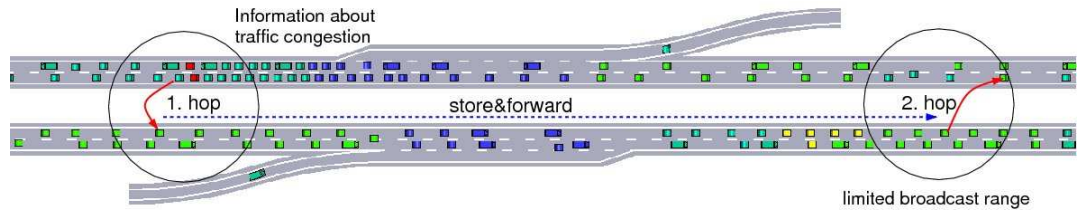}}

 \caption{\label{fig:ivc_hopping}Illustration of the store-and-forward
 strategy using the opposite driving direction for propagating
 messages via short-range inter-vehicle communication in upstream
 direction. First, a message is generated on the occasion of a local
 change in speed. The broadcasted message will be picked up by an
 equipped vehicle in the opposite driving direction (first hop). After
 a certain traveling distance, the vehicle starts broadcasting the
 message which can be received by vehicles in the original driving
 direction (second hop).}
\end{figure}

In the context of freeway traffic, information on the local traffic
situation has to be propagated in an upstream direction. In general,
there are two transport strategies: Either a message ``hops'' from an
equipped car to a subsequent equipped car within the same driving
direction (``longitudinal hopping'') or the message is transmitted to
an IVC-equipped vehicle in the other driving direction which
transports the message upstream and delivers it back by broadcasting
it to cars in the original driving direction (``transversal hopping'',
``cross-transference'' or \textit{store-and-forward}). The latter
strategy is illustrated in Fig.~\ref{fig:ivc_hopping}. Although the
longitudinal hopping process allows for a quasi-instantaneous
information propagation, the connectivity due to the limited
broadcasting range is too weak in the presence of low equipment
rates~\cite{IVC-Goldrain06}. A concept using IVC for traffic-state
detection must therefore tackle the problem that both the required
transport distances into upstream direction and the distances between
two equipped vehicles are typically larger than the broadcasting
range. The transversal hopping mechanism overcomes this problem by
using vehicles in the opposite driving direction as relay
stations. Despite the time delay in receiving messages, the messages
propagate faster than typical shock waves (which are limited to a
speed of \unit[-15]{km/h}, cf.\ Sec.~\ref{sec:wave}).

The microscopic simulation approach is well suited to coupling traffic
and information flows: The movement of vehicles represents a dynamic
network of nodes which determines the spread of  information on the
network. For the purpose of demonstration, let us now simulate the
chain of message propagation by means of IVC in an integrated
simulation:
\begin{enumerate}
\item 
The generation of traffic-related messages by individual vehicles,
\item
the transmission of up-to-date information in upstream direction using
store-and-forward strategy via the opposite driving direction and
\item
the receipt of the messages for predicting the future traffic
situation further downstream.
\end{enumerate}
The object-oriented design of the traffic simulation software (cf.\
Sec.~\ref{sec:simulator}) can be extended in a straightforward way:
First, the simulation of the store-and-forward strategy requires two
independent freeways in opposite directions. Second, each equipped
driver-vehicle agent autonomously detects jam fronts (by means of
velocity gradients) and generates traffic-related messages based on
locally available time series data. To this end, the design of a
vehicle has been extended by a
\texttt{detection} unit which generates traffic-relevant messages
and a \texttt{communication} unit for broadcasting and receiving
messages. Finally, the exchange of messages has been realized by a
\texttt{message pool} which organizes the book-keeping of
message broadcast and reception between equipped cars within a limited
broadcasting range (corresponding to the outdated ether concept). As
the routing in this system is obviously given by the two traffic
streams in opposite directions, no further rules are necessary for
modeling the message exchange process.

\begin{figure}[t]
\centering
  \includegraphics[width=125mm]{\cpath{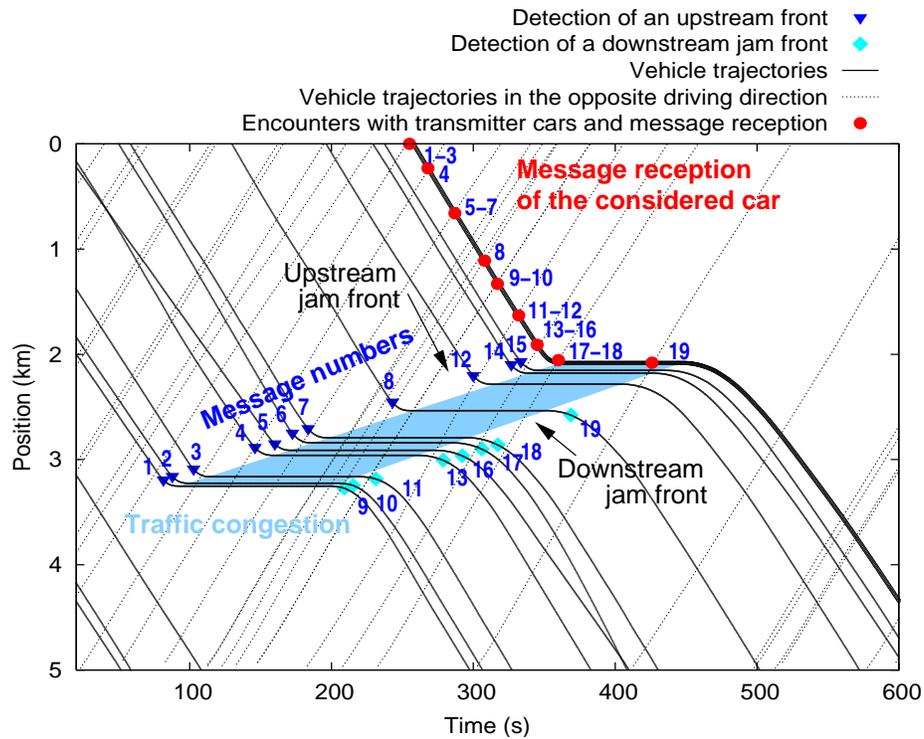}}

 \caption{\label{fig:ivc_stopwave}Space-time diagram of the simulated
 traffic scenario. The trajectories of the IVC-equipped vehicles (3\%)
 are displayed by solid or dotted lines depending on the driving
 direction. The vehicles in the opposite driving direction serve as
 transmitter cars for the store-and-forward strategy.  For the purpose
 of illustration, we have set the maximum broadcasting range to
 \unit[10]{m}. When cars pass the upstream or downstream jam front of
 the moving jam, they broadcast messages (marked by numbers)
 containing the detected position and time.  They are later received
 by the considered vehicle further upstream (thick solid line). Note
 that the crossing trajectories of equipped vehicles (e.g.\ in the
 upper-left corner of the diagram) reflect passing maneuvers due to
 different desired velocities.}
\end{figure}

We consider a scenario with an assumed fraction of only 3\%
communicating vehicles.  The resulting trajectories of equipped
vehicles in both driving directions together with the generation of
messages and their reception by a considered vehicle are illustrated
in Fig.~\ref{fig:ivc_stopwave}. In this scenario, a temporary road
blockage has triggered a stop-and-go wave reflected by horizontal
trajectory curves in one driving direction while the traffic flow in
the opposite driving direction was free. When cars encountered the
propagating stop-and-go wave, they started to broadcast messages about
the detected position and time of the upstream jam front and the
following downstream jam front. The event-driven messages were
received and carried forward by vehicles in the other driving
direction via the store-and-forward mechanism.

As shown in Fig.~\ref{fig:ivc_stopwave}, the considered vehicle
already received the first message about the upcoming traffic
congestion~\unit[2]{km} before reaching the traffic jam. Further
received messages from other equipped vehicles could be used to
confirm and update the upcoming traffic situation further
downstream. Thus, based on a suitable prediction algorithm, each
equipped vehicle could autonomously forecast the moving jam fronts by
extrapolating the spatiotemporal information of the messages. In the
considered simulation scenario, the upstream jam fronts were already
accurately predicted with errors of $\pm \unit[50]{m}$ \unit[1]{km}
ahead of the jam, while the errors for the predicted downstream jam
amounted to $\pm \unit[100]{m}$. Obviously, the quality of the
jam-front anticipation improves with the number and the timeliness of
the incoming messages. More details about the used prediction
algorithm can be found in Ref.~\cite{IVC-Martin-TRR07}.

\section{Conclusions and Future Work}\label{sec:outlook}

Agent-based traffic simulations provide a flexible and customizable
framework for tackling a variety of current research
topics. Simulation of control systems as a part of traffic operations
is an important topic in transport telematics. Due to the
interrelation of the control systems with traffic, both the control
systems and the driver reactions must be described in a combined
simulation framework. Examples are variable message signs and speed
limits, on-ramp metering, lane-changing legislation and dynamic route
guidance systems. An interesting research challenge is adaptive
self-organized traffic control in urban road
networks~\cite{Lammer2007a}.

Furthermore, traffic simulations are used to assess the impacts of
upcoming driver assistance systems such as
\textit{adaptive cruise control} systems on traffic dynamics. The
microscopic modeling approach is most appropriate because it allows
for a natural representation of heterogeneous driver-vehicle agents
and for a detailed specification of the considered models, parameters
and vehicle
proportions~\cite{VanderWerf-TRR-2001,VanderWerf-TRR-2002,ACC-Arne-TRR07,ACC_Hoogendoorn_TRR,Minderhoud,Davis-ACC,tampere2001}. The
challenging question is whether it is possible to design vehicle-based
control strategies aimed at improving the capacity and stability of
traffic flow~\cite{Arne-ACC-TRC}. 

With rapid advances in wireless communication technologies, the
transmission of information within the transportation network is a
challenging issue for the next generation of Intelligent
Transportation Systems (ITS). Agent-based systems form the basis for a
simulation of hybrid systems coupling vehicle and information
flow. The decentralized propagation of information about the upcoming
traffic situation has been discussed as an application for
inter-vehicle communication. Many other applications are conceivable
based on the integration of vehicles and infrastructures implying
vehicle-to-infrastructure communication technologies. However,
realistic and predictive simulations are essential for developing and
testing applications of upcoming communication technologies and
applications.




\newcommand{\etalchar}[1]{$^{#1}$}

\end{document}